\documentclass[12pt, useAMS, usenatbib, nomarkers, referee]{biom}

\makeatletter
\def\endthebibliography{%
  \def\@noitemerr{\@latex@warning{Empty `thebibliography' environment}}%
  \endlist
}
\makeatother

\usepackage{bm}
\usepackage{amssymb,mathtools}
\usepackage{amsmath}
\usepackage{dsfont}
\usepackage{thmtools} 
\usepackage{thm-restate}
\usepackage{booktabs}
\usepackage{adjustbox}
\usepackage{xcolor}
\usepackage{hyperref}
\usepackage{url}
\usepackage{setspace}
\usepackage{multicol}
\usepackage{multirow}
\usepackage{graphicx}
\usepackage{ulem}
\graphicspath{ {./figs/} }
\usepackage{mwe}

\usepackage{algorithm2e}
\usepackage{titlesec}
\titlespacing*{\section}{0pt}{.25\baselineskip}{.1\baselineskip}
\titlespacing*{\subsection}{0pt}{.25\baselineskip}{.1\baselineskip}
\usepackage{calc}
\usepackage{accents}

\usepackage[most]{tcolorbox} 
\usepackage{lipsum}

\tcbset{
  mybox/.style={
    colback=white,
    colframe=black,
    boxrule=0.8pt,
    rounded corners,
    left=6pt,right=6pt,top=6pt,bottom=6pt
  }
}

\tcbset{
  mybox2/.style={
    colback=gray!15,
    colframe=black,
    boxrule=0.8pt,
    rounded corners,
    left=6pt,right=6pt,top=6pt,bottom=6pt
  }
}

\allowdisplaybreaks
\newcommand{\dbtilde}[1]{\accentset{\approx}{#1}}

\title[Nonlinear Coherence for Vector Time Series]{Nonlinear Coherence for Vector Time Series: Defining Region-to-Region Functional Brain Connectivity}

\author{Paolo Victor Redondo$^*$\email{paolovictor.redondo@kaust.edu.sa}, 
Rapha\"el Huser$^{**}$\email{raphael.huser@kaust.edu.sa}, and 
Hernando Ombao$^{***}$\email{hernando.ombao@kaust.edu.sa} \\
King Abdullah University of Science and Technology (KAUST), Thuwal 23955-6900, Saudi Arabia}

\begin{document}

\pagerange{\pageref{firstpage}--\pageref{lastpage}} \pubyear{2025}

\label{firstpage}

\begin{abstract}


Alterations in functional brain connectivity characterize neurodegenerative disorders such as Alzheimer’s disease (AD) and frontotemporal dementia (FTD). As a non-invasive and cost-effective technique, electroencephalography (EEG) is gaining increasing attention for its potential to identify reliable biomarkers for early detection and differential diagnosis of AD and FTD. Considering the behavioral similarities of signals from adjacent EEG channels, we propose a new spectral dependence measure, the \textit{nonlinear vector coherence (NVC)}, to capture beyond-linear interactions between oscillations of two multivariate time series observed from distinct brain regions. This addresses the limitations of conventional channel-to-channel approaches and defines a more natural region-to-region connectivity framework in the frequency domain. As a result, the NVC measure offers a new approach to investigate dependence between brain regions, which then enables to identify altered functional connectivity dynamics associated with AD and FTD. We further introduce a rank-based inference procedure that enables fast and distribution-free estimation of the proposed measure, as well as a fully nonparametric test for spectral independence. The empirical performance of our proposed inference methodology is demonstrated through extensive numerical experiments. An application to resting-state EEG data reveals that our novel NVC measure uncovers distinct and diagnostically meaningful connectivity patterns which effectively discriminate healthy individuals from those with AD and FTD. \\

\end{abstract}


\maketitle

	\section{Introduction}\label{chap:introduction}

Alzheimer's disease (AD) and frontotemporal dementia (FTD) are predominant forms of neurodegenerative disorders that impair cognitive function, and significantly reduce quality of life in affected individuals. AD is primarily characterized by progressive memory loss and cognitive decline \citep{scheltens2021alzheimer}, while FTD leads to behavioral, personality, and language changes due to degeneration of the frontal and temporal lobes \citep{bang2015frontotemporal}. Beyond the financial implications, the emotional and physical toll on patients and healthcare providers underscores the urgency for effective diagnostics and therapeutic strategies for such disorders. Motivated by the substantial burden of these conditions, the objective of this paper is to introduce a new statistical framework that clinicians can use to identify potential alterations in brain functions associated with AD and FTD. Precisely, we develop a novel methodology that enables identification of differences in brain functional networks among healthy individuals, the AD and the FTD population.

\begin{figure}
	\centerline{
		\includegraphics[width=0.75\textwidth]{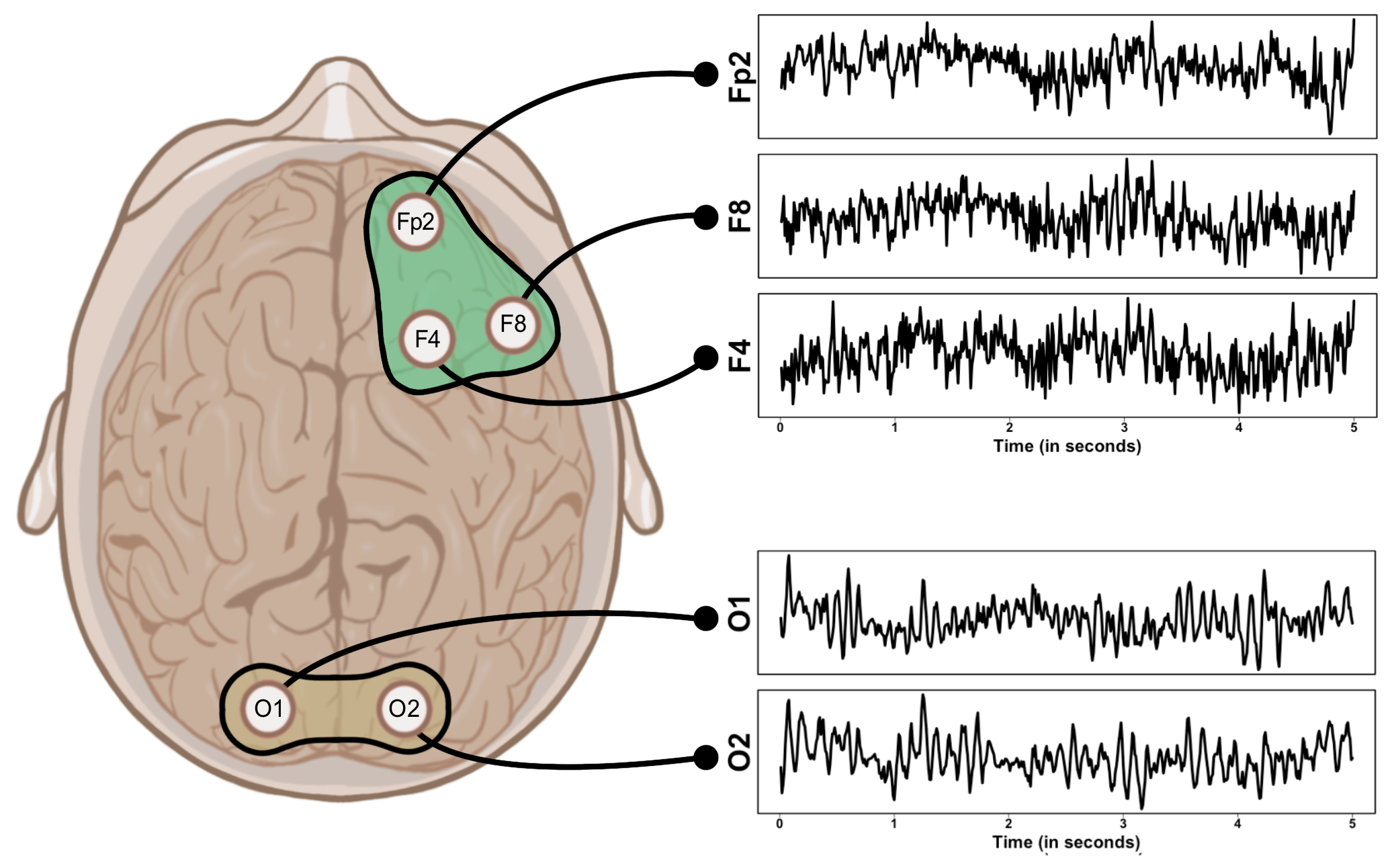}
	}
	\caption{Resting-state EEG recordings from channels Fp2, F4 and F8 in the right frontal region (top) and from channels O1 and O2 in the occipital region (bottom) of a subject with frontotemporal dementia.}
	\label{fig:eeg_recordings}
\end{figure}

Recent advancements have focused on identifying reliable biomarkers, through brain imaging modalities such as electroencephalography (EEG), for the early and accurate diagnosis of AD and FTD \citep{gifford2023biomarkers,modir2023systematic}. As a cost-effective and non-invasive tool, EEG has shown potential in detecting neural alterations associated with these disorders \citep{d2021eeg,tomasello2023neuropsychological}. Specifically, a major research focus is to identify alterations in functional connectivity (i.e., dependence between nodes in a brain network) that distinguish healthy individuals from patients suffering from AD or FTD \citep{wu2024changes}. Standard approaches derive their connectivity metrics between pairs of channels. These, however, have a major drawback since EEG recordings are not well spatially localized due to volume conduction. This is demonstrated in Figure~\ref{fig:eeg_recordings}, where channels in the right frontal region produce highly similar signals and channels from the occipital region also exhibit the same behavior that is distinct from the frontal channels. This suggests that adjacent sensors in a region capture signals from a common neuronal source. Quantifying dependence between signals from a pair of individual channels fails to account for the shared temporal behavior among closely-positioned channels. Hence, it is more appropriate to investigate functional brain connectivity between regions (i.e., groups of spatially adjacent channels), which we pursue in this paper. 

Let $\boldsymbol{X} = (X_1, \ldots, X_p)^{\top}$ and $\boldsymbol{Y} = (Y_1, \ldots, Y_q)^{\top}$ be $p$ and $q$-dimensional random vectors representing the EEG recordings from $p$ and $q$ channels of two distinct brain regions. One way to characterize the dependence between $\boldsymbol{X}$ and $\boldsymbol{Y}$ is through a measure of predictability of $\boldsymbol{Y}$ from $\boldsymbol{X}$, generically denoted as $\kappa(\boldsymbol{Y};\boldsymbol{X})$, which satisfies the following axioms \citep{fuchs2024hierarchical}:
\vspace{-1mm}
\begin{enumerate}
    \item[(A1)] $0 \leq \kappa(\boldsymbol{Y};\boldsymbol{X}) \leq 1$,
    \item[(A2)] $\kappa(\boldsymbol{Y};\boldsymbol{X}) = 0$ if and only if $\boldsymbol{Y}$ and $\boldsymbol{X}$ are independent,
    \item[(A3)] $\kappa(\boldsymbol{Y};\boldsymbol{X}) = 1$ if and only if $\boldsymbol{Y}$ is perfectly dependent on $\boldsymbol{X}$, i.e., there exists some measurable function $\boldsymbol{f} : \mathbb{R}^p \rightarrow \mathbb{R}^q$ such that $\boldsymbol{Y} = \boldsymbol{f}(\boldsymbol{X})$ almost surely.
\end{enumerate}
\noindent For $q = 1$, a specific choice of measure $\kappa$ was introduced by \cite{azadkia2021simple}: the measure $\xi$, defined as
\begin{equation}
    \xi(Y;\boldsymbol{X}) = \frac{\int \mbox{Var}(\mathbb{P}(Y \geq y \mid \boldsymbol{X})) {\rm d} \mu_Y(y)}{\int \mbox{Var}(1_{\{Y \geq y\}}) {\rm d} \mu_Y(y)} \in [0,1],
    \label{eq:xi_AC}
\end{equation}
where $\mu_Y(y)$ is the distribution of $Y$. The $\xi$ measure quantifies the degree of functional (possibly nonlinear) dependence of $Y$ on the variable set $\boldsymbol{X}$ and satisfies the axioms (A1), (A2) and (A3). Subsequently, an extension of $\xi$ to the $q>1$ case was developed by \cite{ansari2025direct}. To be more precise, they defined the measure $T(\boldsymbol{Y};\boldsymbol{X})$ as
\begin{equation}
    T(\boldsymbol{Y};\boldsymbol{X}) = 1 - \frac{q - \sum_{\ell=1}^q \left[\xi(Y_{\ell}~; (\boldsymbol{X}^{\top},Y_{1},\ldots,Y_{\ell-1})^{\top})\right]}{q - \sum_{\ell=1}^q \left[\xi(Y_{\ell}~; (Y_{1},\ldots,Y_{\ell-1})^{\top})\right]},\\
    \label{eq:T_AnF}
\end{equation}
\noindent where, by convention, the expression ``$Y_{1},\ldots,Y_{\ell-1}$'' is taken to be the empty set $\emptyset$ when $\ell = 1$, and $\xi(Y_1~; \emptyset) = 0$. The $T$ measure quantifies the degree of functional dependence of $\boldsymbol{Y}$ on $\boldsymbol{X}$ by sequentially evaluating how each component $Y_{\ell}$ depends on $\boldsymbol{X}$ and on the previously considered components $Y_{1},\ldots,Y_{\ell-1}$. For instance, when $q = 3$, the numerator in Equation~(\ref{eq:T_AnF}) involves the terms $\xi(Y_1;\boldsymbol{X})$, $\xi(Y_2;(\boldsymbol{X}^\top,Y_1)^\top)$ and $\xi(Y_3;(\boldsymbol{X}^\top,Y_1,Y_2)^\top)$. The formula in (\ref{eq:T_AnF}) aggregates these component-wise dependencies into a single measure of predictability that satisfies axioms (A1)--(A3).

Thus, $T(\boldsymbol{Y};\boldsymbol{X})$ offers a promising approach for characterizing complex interactions between groups of signals originating from different brain regions. However, one limitation when applied in the time series context is that it does not identify the frequency oscillations that drive the dependence between $\boldsymbol{X}$ and $\boldsymbol{Y}$. As different frequency oscillations in brain signals are linked to various cognitive interpretations \citep{abhang16bookchp2,redondo2025functional}, this reduces the practical relevance of the $T$ measure in (\ref{eq:T_AnF}) for the analysis of functional brain connectivity. Thus, our solution is formulate a new spectral dependence measure between a pair of vector time series by constructing $T$ in the frequency domain.

\begin{figure}
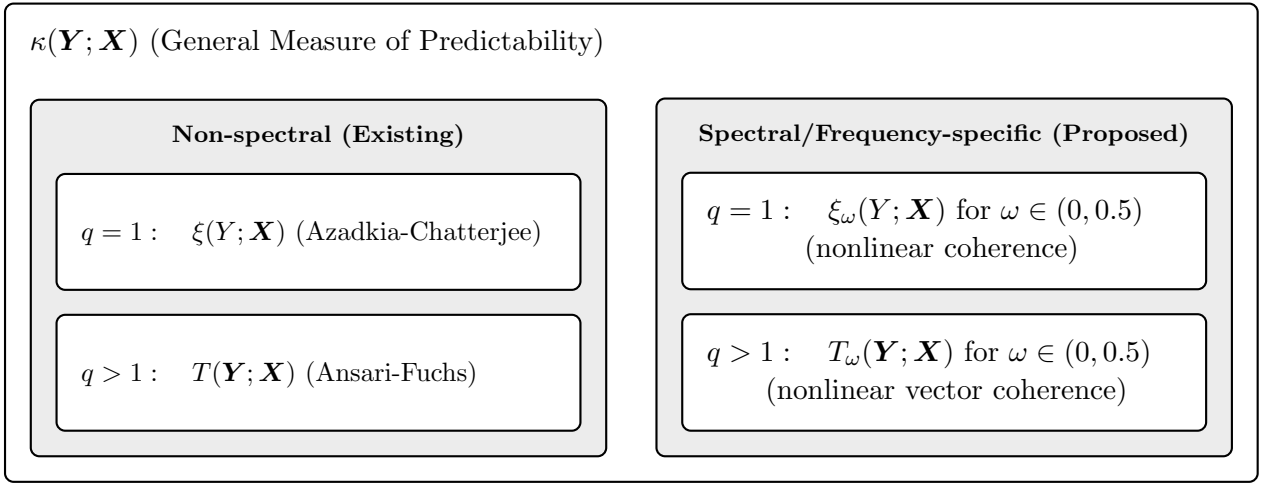

\begin{tcolorbox}[mybox, enhanced]
{\fontsize{11}{12}\selectfont $\kappa(\boldsymbol{Y};\boldsymbol{X})$ (General Measure of Predictability)}
\vspace{5mm}

  \begin{minipage}[t]{0.48\linewidth}
    \begin{tcolorbox}[mybox2]
    {\fontsize{9.5}{12}\selectfont \centering \textbf{Non-spectral~(Existing)}\\}
    
      \begin{tcolorbox}[mybox]  {
      \vspace{0.7em}
      \fontsize{10}{12}\selectfont $q=1:~~~\xi(Y;\boldsymbol{X})$ (Azadkia-Chatterjee)
      \vspace{0.7em}} \end{tcolorbox}\vspace{-0.5em}
      \begin{tcolorbox}[mybox] {
      \vspace{0.7em}
      \fontsize{10}{12}\selectfont $q>1:~~~T(\boldsymbol{Y};\boldsymbol{X})$ (Ansari-Fuchs)
      \vspace{0.7em}} \end{tcolorbox}
    \end{tcolorbox}
  \end{minipage}\hfill
  \begin{minipage}[t]{0.48\linewidth}
    \begin{tcolorbox}[mybox2]
    {\fontsize{9.5}{12}\selectfont \centering \textbf{Spectral/Frequency-specific~(Proposed)}\\}
    
      \begin{tcolorbox}[mybox] {\fontsize{11}{12}\selectfont $q=1:~~~\xi_{\omega}(Y;\boldsymbol{X})$ for $\omega \in (0,0.5)$ 
      
      {~~~~~~~~~~(nonlinear coherence)}} \end{tcolorbox}\vspace{-0.5em}
      \begin{tcolorbox}[mybox] {\fontsize{11}{12}\selectfont $q>1:~~~T_\omega(\boldsymbol{Y};\boldsymbol{X})$ for $\omega \in (0,0.5)$
      
      {~~~~~~(nonlinear vector coherence)}} \end{tcolorbox}
    \end{tcolorbox}
  \end{minipage}
\end{tcolorbox}

    \caption{Schematic diagram of the general measures of predictability containing existing non-spectral dependence metrics and our proposed frequency-specific dependence measures.}
    \label{fig:schematic}
\end{figure}

In this paper, we develop a novel framework for describing cross-region interactions across functional brain networks. Our methodological contributions are summarized as follows.
\begin{enumerate}
    \item We define the new nonlinear vector coherence measure $T_{\omega}(\boldsymbol{Y};\boldsymbol{X})$ to quantify the degree at which the $\omega$-oscillations of a time series $\boldsymbol{Y}=\{\boldsymbol{Y}_t\}$ depend on the $\omega$-oscillations of another time series $\boldsymbol{X}=\{\boldsymbol{X}_t\}$, for $\omega \in (0,0.5)$.
    \item We construct a fully nonparametric estimator for $T_{\omega}(\boldsymbol{Y};\boldsymbol{X})$, which is distribution-free and computable in $O(n \log n)$ time, with $n$ denoting the number of non-overlapping time blocks, thereby enabling efficient inference for the proposed spectral measure.
    \item We develop a nonparametric test for independence in the frequency domain by generating data from the empirical null distribution of the estimator via ``permutations of ranks''.
\end{enumerate}
\noindent Figure~\ref{fig:schematic} illustrates the conceptual framework of the general measure of predictability  $\kappa(\boldsymbol{Y};\boldsymbol{X})$, showing how existing non-spectral measures---$\xi(Y;\boldsymbol{X})$ and $T(\boldsymbol{Y};\boldsymbol{X})$---fit within this framework, and how our proposed spectral measures $\xi_\omega(Y;\boldsymbol{X})$ and $T_\omega(\boldsymbol{Y};\boldsymbol{X})$ generalize these ideas to quantify frequency-specific dependence between multivariate time series.

The remainder of the paper is organized as follows. In Section~\ref{chap:data}, we describe the OpenNeuro public dataset that motivated the development of our work. We then provide the formal definition of the new spectral measure for region-to-region functional brain connectivity in Section~\ref{chap:nlcoherence}, and develop its rank-based estimator and the nonparametric test for spectral independence in Section~\ref{chap:inference}. In Section~\ref{chap:numexp}, we conduct numerical experiments to highlight the attractive properties our inference methodology. In Section~\ref{chap:eeganalysis}, we report interesting results and novel findings on the functional connectivity alterations associated with AD and FTD. Lastly, we conclude and provide future directions of our work in Section~\ref{chap:conclusion}.

	\section{Resting-state EEG Data, Pre-processing and Scientific Problem.}\label{chap:data}

\begin{figure}
	\centerline{
		\includegraphics[width=0.75\textwidth]{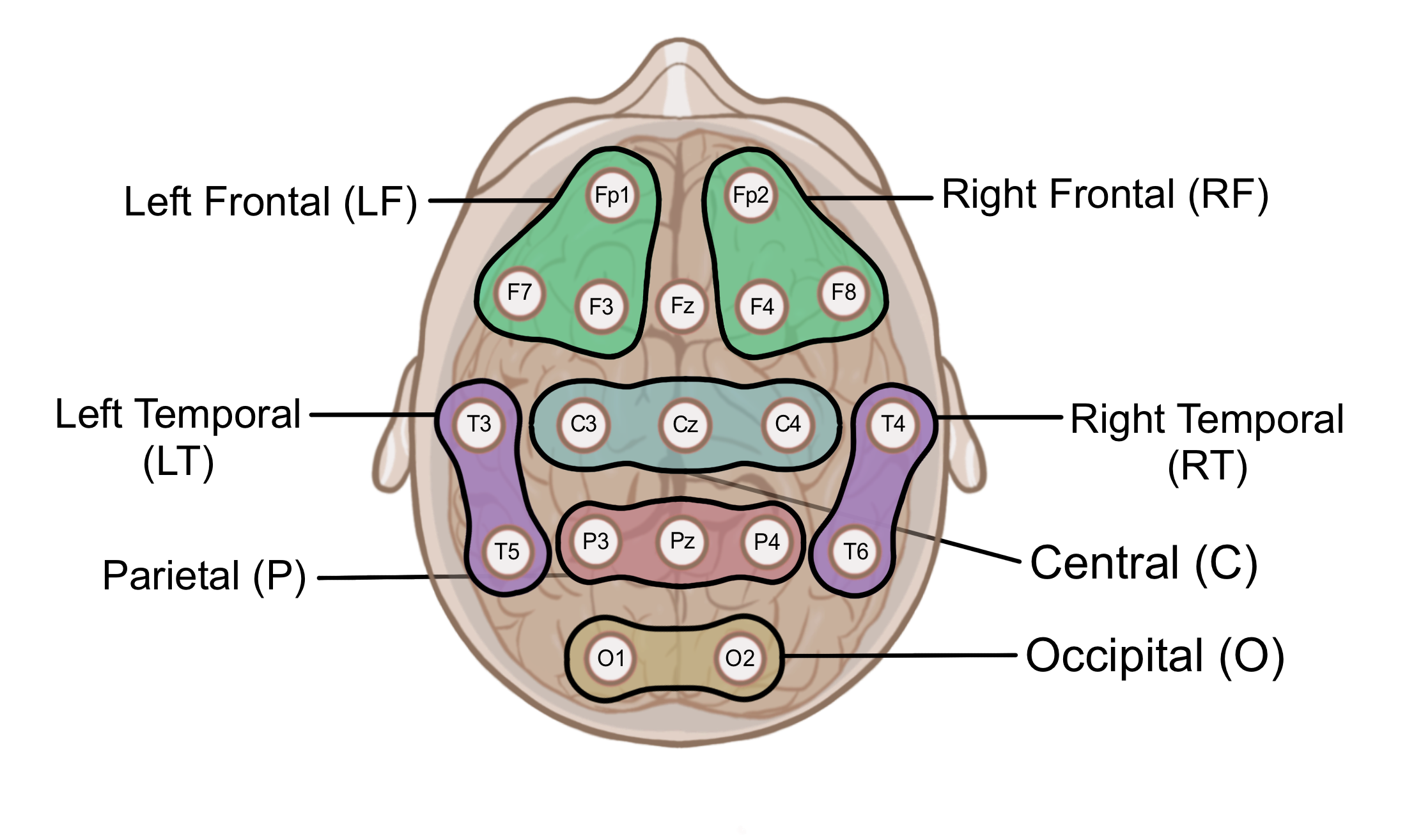}
  }
	\caption{The standard 10--20 19-channel EEG scalp topography and the seven considered regions of interest, namely, the left frontal, right frontal, left temporal, right temporal, central, parietal, and occipital regions.}
	\label{fig:eeg_rois}
\end{figure}

Here, we analyze the OpenNeuro EEG dataset collected by \cite{miltiadous2023dataset}, which is publicly available at \hyperlink{https://openneuro.org/datasets/ds004504}{https://openneuro.org/datasets/ds004504}, and has been analyzed in numerous papers including the works of \cite{miltiadous2023dice}, \cite{wang2024effect} and \cite{nayana2025eeg}. The dataset includes eyes-closed resting-state EEG recordings sampled at 500Hz from 19 channels (see Figure~\ref{fig:eeg_rois}), with two electrodes (A1 and A2) placed on the mastoids serving as the reference electrodes. The recordings were collected from 88 elderly patients (with average age of 66.17 years and standard deviation of 7.36), which consists of 36 subjects diagnosed with AD (AD group), 32 subjects with FTD (FTD group) and 29 healthy controls (CN group). For patients in the disease groups, the median duration of the disease is 25 months with interquantile range being 24--28.5 months. In addition, the cognitive and neuropsychological status of each subject was assessed using the Mini–Mental State Examination (MMSE), a standardized screening tool that ranges from 0 to 30, with lower scores indicating more severe cognitive impairment. The mean (standard deviation) MMSE scores were 17.75 (4.50) for the AD group and 22.17 (8.22) for the FTD group, indicating comparable levels of disease progression within each clinical group. Subjects in the CN group all have MMSE scores of 30, indicating no cognitive decline.

In our analysis, we use the readily available pre-processed EEG signals from each subject, with recording lengths varying from 5.1 to 21.3 minutes. The pre-processing pipeline is outlined as follows. First, a Butterworth band-pass filter was applied to the signals with a frequency range of 0.5 to 45 Hz. Then, the filtered signals were re-referenced to the average value of electrodes A1 and A2. Further, techniques such as the Artifact Subspace Reconstruction (ASR) and Independent Component Analysis (ICA) were implemented to remove persistent or large-amplitude noise and other unwanted artifacts including eye and jaw movements. For exact details of the pre-processing implementations, see \cite{miltiadous2023description}. Lastly, the pre-processed signals were downsampled to 100Hz and were standardized to zero-mean unit variance time series to transform all observations to a unified scale.

As one of our objectives is to investigate alterations in brain functional connectivity that are associated with AD and FTD, we consider exploring dependence between seven brain regions of interests (ROIs), i.e., the left frontal ($LF$), right frontal ($RF$), left temporal ($LT$), right temporal ($RT$), central ($C$), parietal ($P$) and the occipital ($O$) regions. Channels included in each region are illustrated in Figure~\ref{fig:eeg_rois}. Here, we exclude the Fz channel in the analysis to maintain symmetry between the two frontal regions. The choice of these specific regions is justified from previous research (see, e.g., \citealp{dai2025contrastive} and \citealp{sasidharan2025significance}) and the fact that these ROIs are associated with established cognitive functions. For example, the frontal regions are linked to executive and decision-making functions, the temporal regions to recognition of auditory stimuli and memory, the parietal region to spatial orientation and awareness, and the occipital region to visual processing \citep{bjorge2017identification,redondo2025measuring}. Thus, understanding connectivity between the considered ROIs enables a better characterization of neurological dysfunctions related to the two disorders.

Moreover, another interest is the associated impacts of AD and FTD in functional connectivity in the frequency domain. Specifically, we aim to explore dependence alterations among the five canonical frequency bands, namely, the delta ($\delta$; 0.5--4 Hz), theta ($\theta$; 4--8 Hz), alpha ($\alpha$; 8--12 Hz), beta ($\beta$; 12--30 Hz) and gamma ($\gamma$; 30--45 Hz) bands. This is because different oscillations reflect distinct states of the brain. For instance, slow waves such as the delta and theta bands are associated with sleep and drowsiness while fast waves including the alpha, beta and gamma bands are related to attention and concentration \citep{harmony2013functional,abhang16bookchp2,redondo2025functional}. Hence, we develop a novel measure that links the connectivity dysfunctions to the frequency domain, which facilitates neuroscientific interpretations and helps establish their clinical relevance in the development of potential biomarkers for AD and FTD.

In this paper, we differentiate our contributions from other works that have analyzed the same dataset and whose focus is the use of machine learning algorithms to classification of healthy subjects and patients with disorders (see, e.g., \citealp{miltiadous2023dice,kachare2024steadynet} and \citealp{jiang2025classification}). Precisely, our goal instead is to extract features, based on functional connectivity between ROIs, that have high discriminatory power to distinguish between the brain networks of healthy subjects from subjects with AD and FTD. Furthermore, we aim to highlight the shortcomings of current approaches that focus on channel-to-channel dependence, in comparison to our novel approach, which is designed to quantify region-to-region dependence in the frequency domain. We achieve these goals through the application of our proposed methodology in Section~\ref{chap:eeganalysis}.

    
	\section{Novel Measure of Nonlinear Coherence Between Vector Time Series}\label{chap:nlcoherence}

Let $(X,Y)^\top$, with $X = \{X_t\}$ and $Y = \{Y_t\}$, $t=1,2,\ldots$, be a vector time series representing the EEG recordings from two distinct channels. Assuming zero-mean stationarity, these signals can be approximated as a superposition of infinitely many frequency waveforms with random amplitudes \citep{shumwaystoffer17}, i.e.,
\begin{equation}
    \begin{pmatrix} X_t \\ Y_t \end{pmatrix} = \int^{0.5}_{-0.5} \exp(i 2\pi \omega t) \begin{pmatrix} {\rm d}Z_X(\omega) \\ {\rm d}Z_Y(\omega) \end{pmatrix},
    \label{eq:cramer}
\end{equation}
where $i = \sqrt{-1}$, $({\rm d}Z_X(\omega),{\rm d}Z_Y(\omega))^\top$ is a vector of random coefficients from the random increment processes satisfying $\mathbb{E}({\rm d}Z_X(\omega)) = \mathbb{E}({\rm d}Z_Y(\omega)) = 0$ and $\text{Cov}({\rm d}Z_X(\omega),{\rm d}Z_X(\omega')) = \text{Cov}({\rm d}Z_Y(\omega),{\rm d}Z_Y(\omega')) = 0$ for all frequencies $\omega \neq \omega'$, $\omega,\omega' \in (0,0.5)$. Equation~(\ref{eq:cramer}) is commonly known as the Cram{\'e}r representation. Now, let $X^{(\omega)} = \{X^{(\omega)}_t\}$ and $Y^{(\omega)}=\{Y^{(\omega)}_t\}$ be the $\omega$-oscillations of $X$ and $Y$, respectively, where
\begin{equation}
    X^{(\omega)}_t = \exp(i 2\pi \omega t){\rm d}Z_X(\omega) \text{~~~and~~~} Y^{(\omega)}_t = \exp(i 2\pi \omega t){\rm d}Z_Y(\omega).
    \label{eq:omega_XY}
\end{equation}
Coherence between $X$ and $Y$ at frequency $\omega$, denoted by $\rho_{\omega}(X,Y)$, is defined in \cite{ombao2008evolutionary} and in \cite{ombao2024spectral} as
\begin{equation}
    \rho_{\omega}(X,Y) = |\text{Corr}({\rm d}Z_X(\omega),{\rm d}Z_Y(\omega))|^2 = |\text{Corr}(X^{(\omega)},Y^{(\omega)})|^2,
    \label{eq:coh}
\end{equation}
\noindent where for simplicity we have omitted the subscript $t$ in the right-hand side of (\ref{eq:coh}) given that this expression is time-stationary. In other words, coherence is the squared-modulus correlation between the random coefficients or between the $\omega$-oscillations. Suppose $\varphi(U,V)$ is some dependence function between variables $U$ and $V$ and $g_X(\cdot)$, $g_Y(\cdot)$ are some data transformation functions (for which the functional first two moments exist). Expressing coherence as $\rho_{\omega}(X,Y) = \varphi(g_X(X^{(\omega)}),g_Y(Y^{(\omega)}))$, where $\varphi(U,V) = |\text{Corr}(U,V)|^2$ and $g_X(\cdot)$, $g_Y(\cdot)$ are identity functions, we clearly see the limitation of $\rho_{\omega}(X,Y)$. That is, it captures only linear associations between the $\omega$-oscillations as the correlation measure does not account for nonlinear relationships.

Here, we introduce a new spectral measure by considering a more general choice for $\{\varphi,g_X,g_Y\}$. To be more precise, we define the \textit{nonlinear coherence} measure as
\begin{equation}
    \xi_{\omega}(Y;X) = \varphi(g_X(X^{(\omega)}),g_Y(Y^{(\omega)})) = \xi(|Y^{(\omega)}|^2;|X^{(\omega)}|^2),
    \label{eq:nlcoh}
\end{equation}
where, similarly as above, we omit the subscript $t$ in (\ref{eq:nlcoh}) and throughout the rest of this section because the squared modulus of $X^{(\omega)}_t$ and $Y^{(\omega)}_t$ is time-stationary. In other words, we take $\varphi(U,V)$ to be $\xi(V;U)$ defined in Equation~(\ref{eq:xi_AC}) and $g_X(\cdot)$, $g_Y(\cdot)$ to be the modulus squared function. The advantage of our proposed spectral measure $\xi_{\omega}(Y;X)$ over $\rho_{\omega}(X,Y)$ is that the former captures general types of nonlinear dependence between the $\omega$-oscillations. Furthermore, $\xi_{\omega}(Y;X)$ inherits all the nice properties of $\xi$, e.g., it is a measure of predictability in the frequency domain, which means that it satisfies axioms (A1)--(A3).

Suppose now we want to quantify the spectral dependence between two groups of signals originating from different regions in the brain, which is a common interest among neuroscientists. Denote by $\boldsymbol{X} = (X_{1},\ldots,X_{p})^\top$ and $\boldsymbol{Y} = (Y_{1},\ldots,Y_{q})^\top$ the EEG recordings from two brain regions consisting of $p$ and $q$ channels, respectively, and let their $\omega$-oscillations be $\boldsymbol{X}^{(\omega)} = (X^{(\omega)}_{1},\ldots,X^{(\omega)}_{p})^\top$ and $\boldsymbol{Y}^{(\omega)} = (Y^{(\omega)}_{1},\ldots,Y^{(\omega)}_{q})^\top$, as defined in Equation~(\ref{eq:omega_XY}). Moreover, consider $\widetilde{\boldsymbol{X}}^{(\omega)} = (\widetilde{X}^{(\omega)}_{1},\ldots,\widetilde{X}^{(\omega)}_{p})^\top \coloneq (|X^{(\omega)}_{1}|^2,\ldots,|X^{(\omega)}_{p}|^2)^\top$ and $\widetilde{\boldsymbol{Y}}^{(\omega)} = (\widetilde{Y}^{(\omega)}_{1},\ldots,\widetilde{Y}^{(\omega)}_{q})^\top \coloneq (|Y^{(\omega)}_{1}|^2,\ldots,|Y^{(\omega)}_{q}|^2)^\top$. We define the \textit{nonlinear vector coherence (NVC)} measure to be
\begin{equation}
    \begin{aligned}
        T_{\omega}(\boldsymbol{Y};\boldsymbol{X}) \coloneq &~~ T(\widetilde{\boldsymbol{Y}}^{(\omega)};\widetilde{\boldsymbol{X}}^{(\omega)})\\
        =&~~1 - \frac{q - \sum_{\ell=1}^q \left[\xi(\widetilde{Y}^{(\omega)}_{\ell}~; (\widetilde{\boldsymbol{X}}^{(\omega)\top},\widetilde{Y}^{(\omega)}_{1},\ldots,\widetilde{Y}^{(\omega)}_{\ell-1})^{\top})\right]}{q - \sum_{\ell=1}^q \left[\xi(\widetilde{Y}^{(\omega)}_{\ell}~; (\widetilde{Y}^{(\omega)}_{1},\ldots,\widetilde{Y}^{(\omega)}_{\ell-1})^{\top})\right]}, \text{~with~} \xi(\widetilde{Y}^{(\omega)}_{1}; \emptyset) = 0,
        \label{eq:nlcoh_vec}
    \end{aligned}    
\end{equation}
\noindent where the $T$ and $\xi$ measures are those defined in (\ref{eq:xi_AC}) and (\ref{eq:T_AnF}), respectively, and the expression ``$\widetilde{Y}^{(\omega)}_{1},\ldots,\widetilde{Y}^{(\omega)}_{\ell-1}$'' is taken to be empty when $\ell = 1$.

As a natural formulation of the $T$ measure in the frequency domain, we emphasize that $T_{\omega}(\boldsymbol{Y};\boldsymbol{X})$ measures the nonlinear spectral interactions between the $\omega$-oscillations of $\boldsymbol{X}$ and $\boldsymbol{Y}$, and hence, it captures the functional connectivity between the brain regions associated with the signals $X_{1},\ldots,X_{p}$ and $Y_{1},\ldots,Y_{q}$. More importantly, $T_{\omega}(\boldsymbol{Y};\boldsymbol{X}) = 0$ is equivalent to the independence of $\boldsymbol{Y}$ from $\boldsymbol{X}$ at frequency $\omega$. Another advantage of our new measure is that $T_{\omega}(\boldsymbol{Y};\boldsymbol{X}) \in [0,1]$, which leads to straightforward interpretations and facilitate comparisons of magnitudes of spectral dependence between several groups of signals. These appealing properties make the NVC measure a novel addition to the statistical toolbox that may be used by neurologists and clinicians to explore brain connectivity.

It is important to note, however, that because the $\xi$ measure is not symmetric, i.e., $\xi(U_1;U_2) \neq \xi(U_2;U_1)$ in general, for variables $U_1$ and $U_2$, the NVC measure $T_{\omega}(\boldsymbol{Y};\boldsymbol{X})$ is sensitive to the permutations of the components of $\widetilde{\boldsymbol{Y}}^{(\omega)}$. To overcome this limitation, we consider a permutation-invariant analogue of $T_{\omega}$ defined as
\begin{equation}
    \Bar{T}_{\omega}(\boldsymbol{Y};\boldsymbol{X}) \coloneq \frac{1}{q!} \sum_{\boldsymbol{\sigma} \in S_q} T(\widetilde{\boldsymbol{Y}}^{(\omega)}_{\boldsymbol{\sigma}};\widetilde{\boldsymbol{X}}^{(\omega)}) \coloneq \mathbb{E}_{S_q}\left(T(\widetilde{\boldsymbol{Y}}^{(\omega)}_{\boldsymbol{\sigma}};\widetilde{\boldsymbol{X}}^{(\omega)})\right),
    \label{eq:nlcoh_vec_PI}
\end{equation}
where $S_q$ denote the set of all possible permutations of $\{1,\ldots,q\}$ and $\widetilde{\boldsymbol{Y}}^{(\omega)}_{\boldsymbol{\sigma}} = (\widetilde{Y}^{(\omega)}_{\sigma_1},\ldots,\widetilde{Y}^{(\omega)}_{\sigma_q})^\top$ for $\boldsymbol{\sigma} = (\sigma_1,\ldots,\sigma_q)^\top \in S_q$, while $\mathbb{E}_{S_q}(\cdot)$ denotes the expectation taken over the uniform distribution over $S_q$. Moreover, a symmetric permutation-invariant measure may be obtained by considering $\Bar{T}^*_{\omega}(\boldsymbol{Y},\boldsymbol{X}) = \max \{\Bar{T}_{\omega}(\boldsymbol{Y};\boldsymbol{X}),\Bar{T}_{\omega}(\boldsymbol{X};\boldsymbol{Y})\}$. Clearly, $\Bar{T}_{\omega}(\boldsymbol{Y};\boldsymbol{X})$ and $\Bar{T}^*_{\omega}(\boldsymbol{Y},\boldsymbol{X})$ inherits the properties of $T_{\omega}(\boldsymbol{Y};\boldsymbol{X})$, and thus, can be shown to be measures of predictability in the frequency domain.

In addition, one of our objective is to provide empirical evidence of the advantages of our proposed NVC measure over standard channel-to-channel connectivity approaches. In particular, one such approach is the so-called ``band coherence", which is defined as follows. Let $X^{(\Omega)}$ and $Y^{(\Omega)}$ be the $\Omega$-band oscillations from two channels where $\Omega \subset (0,0.5)$ is a set of individual frequencies. Following \cite{ombao2024spectral}, the pairwise band coherence (PBC) between $X^{(\Omega)}$ and $Y^{(\Omega)}$ is expressed as 
\begin{equation}
    \text{PBC}_{\Omega}(X,Y) = |\text{Corr}(X^{(\Omega)},Y^{(\Omega}))|^2,
    \label{eq:PBC}
\end{equation}
i.e., the squared correlation between the two band-specific time series. The measure (\ref{eq:PBC}), defined over ``frequency band''-specific oscillations, is analogous to the coherence measure in Equation~(\ref{eq:coh}), which is defined for specific frequency $\omega \in (0,0.5)$. We emphasize that the PBC measure is formulated only for a pair of individual channels and not between a pair of groups of channels. Thus, when capturing dependence between brain regions via PBC, a common strategy is to aggregate (e.g., by simple averaging) the PBC values across all possible channel pairs within ROIs. The use of such arbitrary aggregation may lead to inappropriate characterization of the dependence structure between brain regions, which is naturally addressed by our novel NVC measure. We demonstrate and discuss this in our data analysis in Section~\ref{chap:eeganalysis}.


	\section{Inference Methodology for Region-to-Region Functional Brain Connectivity}\label{chap:inference}

In this section, we detail the steps to perform inference using our novel NVC measure. This includes the nonparametric estimation of our measure, which involves the integration of periodograms in a rank-based estimation scheme (see Section~\ref{subchap:estimation}). Then, we introduce a nonparametric test for independence in the frequency domain, through a new strategy which we call the ``permutation of ranks", in Section~\ref{subchap:nonpartest}.

\subsection{Notation and Conventions}\label{subchap:notation}

All dependence measures described in the previous sections are treated as functions of random variables, whereas their corresponding estimators, which we develop in this section, are computed from collections of realizations of these variables. For example, $\xi(Y; \boldsymbol{X})$ in Equation~(\ref{eq:xi_AC}), represents a population-level dependence measure between the random variable $Y$ and the random vector $\boldsymbol{X}$. Its sample analogue, denoted by $\xi_n(Y; \boldsymbol{X})$, is computed from $n$ independent and identically distributed (i.i.d.) observations $\{Y_j, \boldsymbol{X}_j\}$, $j=1,\ldots,n$. For notational simplicity, we express all estimators as functions of the random variables with the subscript $n$, thereby implicitly indicating that they are based on the corresponding observed realizations.

\subsection{Estimation of the Nonlinear Vector Coherence Measure}\label{subchap:estimation}

Recall that the $\ell$-th component of $\widetilde{\boldsymbol{Y}}^{(\omega)}$, denoted by $\widetilde{Y}^{(\omega)}_{\ell}$, is given, at a fixed time point $t$, by $\widetilde{Y}^{(\omega)}_{\ell,t} = |Y^{(\omega)}_{\ell,t}|^2 = |\exp(i 2\pi \omega t){\rm d}Z_{Y_{\ell}}(\omega)|^2 = |{\rm d}Z_{Y_{\ell}}(\omega)|^2$, where $\ell = 1, \ldots, q$. Similarly, the $\ell'$-th component of $\widetilde{\boldsymbol{X}}^{(\omega)}$, denoted by $\widetilde{X}^{(\omega)}_{\ell'}$ for $\ell'=1,\ldots,p$, is defined as $\widetilde{X}^{(\omega)}_{\ell',t} = |{\rm d}Z_{X_{\ell'}}(\omega)|^2$, given a fixed $t$. Since these random coefficients are unobservable latent components of the recorded signals, estimating $T_{\omega}$ requires finding suitable data analogues for $|{\rm d}Z_{Y_{\ell}}(\omega)|^2$ and $|{\rm d}Z_{X_{\ell'}}(\omega)|^2$. Our solution involves the use of periodograms calculated over non-overlapping time segments. Specifically, we divide the observed signals into $n$ non-overlapping blocks of length $B$, and~denote the $j$-th block as $\{(Y_{1,t},\ldots,Y_{q,t})^\top\}_{t \in \tau_j}$ and $\{(X_{1,t},\ldots,X_{p,t})^\top\}_{t \in \tau_j}$, where $\tau_j~=~\{B(j-1)+1,\ldots,Bj\}$, $j = 1, \ldots, n$. Then, the discrete Fourier transform of the $\ell$-th component in the $j$-th block of $\boldsymbol{Y}$ is $d^{j}_{Y_{\ell}}(\omega_k) = B^{-\frac{1}{2}} \sum^B_{t = 1} Y_{\ell,B(j-1)+t} \exp(-i 2\pi \omega_k t)$, evaluated at the Fourier frequencies $\omega_k = k/B$, for $k = 0, 1, \ldots, B-1$, and the corresponding periodogram is then $I^{j}_{Y_{\ell}}(\omega_k) = |d^{j}_{Y_{\ell}}(\omega_k)|^2$. Intuitively, for $B \rightarrow \infty$, one has $\omega_k \rightarrow \omega \neq 0$, and $\mathbb{E}(I^{j}_{Y_{\ell}}(\omega_k)) \rightarrow f_{Y_{\ell}}(\omega)$, where $f_{Y_{\ell}}(\omega)$ is the spectral density of $Y_{\ell}$ (see \citealp{shumwaystoffer17} for the derivations) with $\mathbb{E}(|{\rm d}Z_{Y_{\ell}}(\omega)|^2) = \text{Var}({\rm d}Z_{Y_{\ell}}(\omega)) = f_{Y_{\ell}}(\omega){\rm d}\omega$. The corresponding periodogram of the $\ell'$-th component in the $j$-th block of $\boldsymbol{X}$, denoted by $I^{j}_{X_{\ell'}}(\omega_k)$, is obtained in a similar manner. Here, we treat the calculated component-wise periodograms across all time blocks, i.e., $\{I^{j}_{Y_{\ell}}(\omega_k)\}$ and $\{I^{j}_{X_{\ell'}}(\omega_k)\}$ with $j = 1, \ldots, n$, as realizations of some random variables $I_{Y_{\ell}}(\omega_k)$ and $I_{X_{\ell'}}(\omega_k)$ associated with the frequency $\omega_k$. These realizations therefore serve as appropriate data analogues for $|{\rm d}Z_{Y_{\ell}}(\omega)|^2$ and $|{\rm d}Z_{X_{\ell'}}(\omega)|^2$. Hence, we propose to estimate nonlinear vector coherence using $\dbtilde{\boldsymbol{Y}}^{(\omega_k)} = (\dbtilde{Y}^{(\omega_k)}_1, \ldots, \dbtilde{Y}^{(\omega_k)}_q)^\top \coloneq (I_{Y_{1}}(\omega_k),\ldots,I_{Y_{q}}(\omega_k))^\top$ and $\dbtilde{\boldsymbol{X}}^{(\omega_k)} = (\dbtilde{X}^{(\omega_k)}_1, \ldots, \dbtilde{X}^{(\omega_k)}_p)^\top \coloneq (I_{X_{1}}(\omega_k),\ldots,I_{X_{p}}(\omega_k))^\top$, observed across all $n$ non-overlapping time blocks. 

Precisely, for $\omega_k = k/B, k = 0, 1, \ldots, B-1$, our rank-based estimator for $T_{\omega}(\boldsymbol{Y};\boldsymbol{X})$ is given by
\begin{equation}
    \begin{aligned} 
        T_{\omega_k,n}(\boldsymbol{Y};\boldsymbol{X}) \coloneq & ~~1 - \frac{q - \sum_{\ell=1}^q \left[\xi_n(\dbtilde{Y}^{(\omega_k)}_{\ell}~; (\dbtilde{\boldsymbol{X}}^{(\omega_k)\top},\dbtilde{Y}^{(\omega_k)}_{1},\ldots,\dbtilde{Y}^{(\omega_k)}_{\ell-1})^{\top})\right]}{q - \sum_{\ell=2}^q \left[\xi_n(\dbtilde{Y}^{(\omega_k)}_{\ell}~; (\dbtilde{Y}^{(\omega_k)}_{1},\ldots,\dbtilde{Y}^{(\omega_k)}_{\ell-1})^{\top})\right]}, 
        \label{eq:T_omega_n}
    \end{aligned}
\end{equation}
where ``$\dbtilde{Y}^{(\omega_k)}_{1},\ldots,\dbtilde{Y}^{(\omega_k)}_{\ell-1}$'' is taken as the empty set when $\ell=1$, and for a random variable $U$ and a random vector $\boldsymbol{V}$ with $n$ i.i.d. realizations $\{U_j\}$ and $\{\boldsymbol{V}_j\}$, $j=1,\ldots,n$,
\begin{align}
     \xi_n(U;\boldsymbol{V}) \coloneq& ~\frac{\sum^{n}_{j=1} \left[n \min \{R_j,R_{N(j)}\} - L^2_j\right]}{\sum^{n}_{j=1} L_j (n - L_j)}.
    \label{eq:AzadChat}
\end{align}
Here, $R_j$ denotes the rank of $U_j$ among $U_1, \ldots, U_n$ and $L_j$ denotes the number of indices $j'\in\{1,\ldots,n\}$ such that $U_{j'} \leq U_j$. The index $N(j)$ represents the element $j'~\in~\{1,\ldots,n\}$ such that $\boldsymbol{V}_{j'}$ is the nearest neighbor of $\boldsymbol{V}_j$ with respect to the Euclidean distance (and with ties broken at random if they exist). The statistic $T_{\omega_k,n}$ in Equation~(\ref{eq:T_omega_n}) is inspired by the estimator of \cite{ansari2025direct}, but is tailored to capturing functional connectivity between brain regions in the frequency domain. Although dependent on the choice of block length $B$, which is the only tuning parameter to compute $T_{\omega_k,n}$, our proposed estimator has nice properties, including that it is fully nonparametric and allows for fast computations since $\xi_n$ in Equation~(\ref{eq:AzadChat}) can be computed in $O(n \log n)$ time. Moreover, the estimators of the permutation-invariant metric $\Bar{T}_{\omega}$ and the symmetric permutation-invariant $\Bar{T}^*_{\omega}$, denoted by $\Bar{T}_{\omega_k,n}$ and $\Bar{T}^*_{\omega_k,n}$, respectively, may be defined based on $T_{\omega_k,n}$. Hence, the contribution of our work is a new fast-to-calculate distribution-free spectral dependence metric that enables for investigating region-to-region functional brain connectivity.

A caveat, however, is that when dealing with large-dimensional vector time series, calculation of the permutation-invariant metric $\Bar{T}_{\omega_k,n}(\boldsymbol{Y};\boldsymbol{X})$ may be computationally challenging. Specifically, as $q$ increases, the total number of possible permutations required to calculate $\Bar{T}_{\omega_k,n}$ becomes impractical as it involves computations of order $q!$. Such a case is typically encountered when analyzing data from other imaging modalities such as functional magnetic resonance imaging (fMRI), magnetoencephalograms and local field potentials, where for example, in fMRI, brain regions are commonly defined over a large number of voxels.

One natural way to address this challenge is to consider only a sufficient number of random permutations, say $Q$ such that $Q \ll q!$, instead of exhausting all possible $q!$ permutations, when calculating $\Bar{T}_{\omega_k,n}$. Explicitly, we define the permutation-invariant estimator $\Bar{T}_{\omega_k,n}$ as
\begin{equation}
    \Bar{T}_{\omega_k,n} = \frac{1}{Q} \sum_{\boldsymbol{\sigma} \in S^{'}_q} T_{\omega_k,n}(\boldsymbol{Y}_{\boldsymbol{\sigma}};\boldsymbol{X}),
    \label{eq:T_omega_n_PI}
\end{equation}
where $S^{'}_q$ denote a set containing $Q$ random permutations of $\{1,\ldots,q\}$ drawn uniformly without replacement from $S_q$, and $\boldsymbol{Y}_{\boldsymbol{\sigma}} = (Y_{\sigma_1},\ldots,Y_{\sigma_q})^\top$, which implicitly involves the permuted component-wise periodograms $\dbtilde{\boldsymbol{Y}}^{(\omega_k)}_{\boldsymbol{\sigma}} = (\dbtilde{Y}^{(\omega_k)}_{\sigma_1},\ldots,\dbtilde{Y}^{(\omega_k)}_{\sigma_q})^\top$ at frequency $\omega_k$ with $\dbtilde{Y}^{(\omega_k)}_{\ell}$ as defined in Equation~(\ref{eq:T_omega_n}), for $\boldsymbol{\sigma} = (\sigma_1,\ldots,\sigma_q)^\top \in S^{'}_q$. In the Supplementary Material, we show through numerical experiments that, for $\omega_k \approx \omega$, $\Bar{T}_{\omega_k,n}$ is indeed as unbiased estimator of $\Bar{T}_{\omega}$.

\subsection{Nonparametric Test for Spectral Independence}\label{subchap:nonpartest}

One appealing feature of $T_{\omega}(\boldsymbol{Y};\boldsymbol{X})$ is that $T_{\omega}(\boldsymbol{Y};\boldsymbol{X}) = 0$ if and only if the $\omega$-oscillations of the time series $\boldsymbol{X}=\{\boldsymbol{X}_t\}$ and $\boldsymbol{Y}=\{\boldsymbol{Y}_t\}$ are independent, which we describe as \textit{spectral independence} at frequency $\omega$. In the finite data setting, one may consider constructing a confidence interval based on $T_{\omega_k,n}$ and conclude independence at frequency $\omega$ when zero is inside the calculated interval. However, such approach is quite challenging. Precisely, even when assuming asymptotic normality to hold for $T_{\omega_k,n}$, estimating its limiting variance (which is generally unknown) is not straightforward. This causes problems as constructing confidence intervals based on a poorly estimated asymptotic variance may lead to a false characterization of dependence between the oscillations.

Our solution instead is to develop a nonparametric test for independence in the frequency domain where we obtain the empirical null distribution of our estimator $T_{\omega_k,n}$ via ``permutation of ranks''. Specifically, suppose we wish to test the null hypothesis $H_0: T_{\omega}(\boldsymbol{Y};\boldsymbol{X}) = 0$, i.e., $\boldsymbol{X}$ and $\boldsymbol{Y}$ are independent at frequency $\omega$, against $H_1: T_{\omega}(\boldsymbol{Y};\boldsymbol{X}) \neq 0$, i.e., $\boldsymbol{X}$ and $\boldsymbol{Y}$ are dependent at frequency $\omega$. For a given dataset and choice of block length $B$, consider the proposed estimator $T_{\omega_k,n}$ defined in Equation~(\ref{eq:T_omega_n}). This expression involves the quantities $\xi_n(U^{(\omega_k)}_{\ell};\boldsymbol{V}^{(\omega_k)}_{\ell})$ (in the numerator) and $\xi_n(U^{(\omega_k)}_{\ell};\boldsymbol{V}^{(\omega_k)}_{-\boldsymbol{X},\ell})$ (in the denominator), for $\ell = 1, \ldots, q$, where $U^{(\omega_k)}_{\ell} = \dbtilde{Y}^{(\omega_k)}_{\ell}$, $\boldsymbol{V}^{(\omega_k)}_{\ell} = (\dbtilde{\boldsymbol{X}}^{(\omega_k)\top},\dbtilde{Y}^{(\omega_k)}_{\ell-1},\ldots,\dbtilde{Y}^{(\omega_k)}_{1})^\top$, and $\boldsymbol{V}^{(\omega_k)}_{-\boldsymbol{X},\ell} = (\dbtilde{Y}^{(\omega_k)}_{\ell-1},\ldots,\dbtilde{Y}^{(\omega_k)}_{1})^\top$, which consist of the component-wise periodograms of $\boldsymbol{X}$ and $\boldsymbol{Y}$. Since $T_{\omega}(\boldsymbol{Y};\boldsymbol{X}) = 0$ implies that $\xi(U^{(\omega)}_{\ell};\boldsymbol{V}^{(\omega)}_{\ell}) = \xi(U^{(\omega)}_{\ell};\boldsymbol{V}^{(\omega)}_{-\boldsymbol{X},\ell})$ for all $\ell$, we construct the empirical null distribution of $T_{\omega_k,n}$ based on the behavior of $\xi_{n}$ under independence by focusing on the case when $\xi(U^{(\omega)}_{\ell};\boldsymbol{V}^{(\omega)}_{\ell}) = \xi(U^{(\omega)}_{\ell};\boldsymbol{V}^{(\omega)}_{-\boldsymbol{X},\ell}) = 0$.

It is straightforward to show that, for any random variable $U$ and random vector $\boldsymbol{V}$, with $n$ i.i.d. samples $\{U_j\}$ and $\{\boldsymbol{V}_j\}$, $j=1,\ldots,n$, the statistic $\xi_{n}(U;\boldsymbol{V}) \rightarrow 1$ when the ranks $R_j$ of $U_j$ are close to the corresponding ranks $R_{N(j)}$ for all $j=1,\ldots,n$. That is, $\xi_n$ captures strong functional dependence of $U$ on $\boldsymbol{V}$ when the rank of the $j$-th observation of $U$ is close to the rank of its nearest neighbor based on $\boldsymbol{V}$. In contrast, $\xi_{n}(U;\boldsymbol{V}) \rightarrow 0$ when $R_j$ and $R_{N(j)}$ are highly dissimilar for all $j=1,\ldots,n$, i.e., if the ranks $R_{N(j)}$ resemble a ``shuffled'' version of the ranks $R_j$, independent of $\boldsymbol{V}$. Therefore, we can generate the empirical null distribution of $\xi_n$, and thus also the distribution of $T_{\omega_k,n}$ under $H_0$, by replacing $R_j$ with random permutations of $\{1,\ldots,n\}$ and $R_{N(j)}$ with random samples (with replacement) of size $n$ from $\{1,\ldots,n\}$ in the estimation.

The outline of our proposed nonparametric test for spectral independence between groups of signals is detailed as follows:
\vspace{-5mm}
\begin{enumerate}
    \item Given the observed data, divided into $n$ non-overlapping time segments of length $B$, denote the original NVC estimates between the two vector series~by
    \begin{equation*}
        T_{\omega_k,n}(\boldsymbol{Y};\boldsymbol{X}) \coloneq 1 - \frac{q - \sum_{\ell=1}^q \left[\xi_{n}(U^{(\omega_k)}_{\ell};\boldsymbol{V}^{(\omega_k)}_{\ell})\right]}{q - \sum_{\ell=2}^q \left[\xi_{n}(U^{(\omega_k)}_{\ell};\boldsymbol{V}^{(\omega_k)}_{-\boldsymbol{X},\ell})\right]}, ~~~\text{for}~\omega_k = \frac{k}{B}, k = 0, 1, \ldots, B-1.
    \end{equation*}

    \item Generate a random permutation of $\{1,\ldots,n\}$ and denote it by $R^{0}_j$. Subsequently, define $L^{0}_j$ as $n-R^{0}_j+1$. Then, draw a random sample with replacement of size $n$ from $\{1,\ldots,n\}$ and denote it by $R^{0}_{N(j)}$. Compute $\xi_n$, recall Equation~(\ref{eq:AzadChat}), by replacing the ranks $R_j$, $L_j$ and $R_{N(j)}$ with $R^{0}_j$, $L^{0}_j$ and $R^{0}_{N(j)}$, respectively, and denote the estimate by $\xi^{0}_n$. 

    \item Compute $T_{\omega_k,n}(\boldsymbol{Y};\boldsymbol{X})$ under $H_0$, by replacing $\xi_{n}(U^{(\omega_k)}_{\ell};\boldsymbol{V}^{(\omega_k)}_{\ell})$ and $\xi_{n}(U^{(\omega_k)}_{\ell};\boldsymbol{V}^{(\omega_k)}_{-\boldsymbol{X},\ell})$,\\ 
    $\ell = 1, \ldots, q$, each with a random realization of $\xi^{0}_n$, and repeat this procedure $R$ times to get the estimates $T^{(r)}_{\omega_k,n}(\boldsymbol{Y};\boldsymbol{X}),$ $r=1,\ldots,R$.

    \item For each $\omega_k$, compute the p-value associated with $T_{\omega_k,n}(\boldsymbol{Y};\boldsymbol{X})$ as the relative frequency of the event $\{T^{(r)}_{\omega_k,n}(\boldsymbol{Y};\boldsymbol{X}) \geq T_{\omega_k,n}(\boldsymbol{Y};\boldsymbol{X})\}$ over the $R$ replicates.

    \item Given a specified nominal level $\alpha \in (0,1)$, reject the null hypothesis if the p-value is less than $\alpha$. Otherwise, do not reject $H_0$.
\end{enumerate}
\vspace{-5mm}
\noindent Tests based on $\Bar{T}_{\omega_k,n}(\boldsymbol{Y};\boldsymbol{X})$ and $\Bar{T}^*_{\omega_k,n}(\boldsymbol{Y},\boldsymbol{X})$, i.e., the permutation-invariant and the symmetric permutation-invariant versions of the NVC measure, can be constructed in a similar manner. 

One advantage of the outlined approach is that the same strategy of considering random permutations of ranks may be employed to construct the empirical null distribution for the $T_n$ estimator of \cite{ansari2025direct}, which then corresponds to a test for independence, although for general random variables. Another benefit of our nonparametric test is that the empirical null distribution of $T_{\omega_k,n}(\boldsymbol{Y};\boldsymbol{X})$, given by $\{T^{(r)}_{\omega_k,n}(\boldsymbol{Y};\boldsymbol{X})\}$, $r=1,\ldots,R$, is independent of the actual values in the observed data. In fact, the distribution of $T^{(r)}_{\omega_k,n}(\boldsymbol{Y};\boldsymbol{X})$ only depends on the number of observations $n$ and the dimension $q$, which may be interpreted as the ``degrees of freedom" of the reference distribution. Moreover, $T^{(r)}_{\omega_k,n}(\boldsymbol{Y};\boldsymbol{X})$ does not depend on the frequency $\omega_k$. This is because the calculation of $T^{(r)}_{\omega_k,n}(\boldsymbol{Y};\boldsymbol{X})$ only requires random permutations and random samples (with replacement) of $\{1,\ldots,n\}$, which is independent from the frequency $\omega_k$. As a result, generating p-values for $T_{\omega_k,n}(\boldsymbol{Y};\boldsymbol{X})$ for all $\omega_k$ only requires a single generation of the empirical null distribution, which depends solely on $n$ and $q$. This renders the proposed approach computationally efficient. Thus, we offer a novel procedure to test for independence between vector time series in the frequency domain that involves fast computations, even when considering large number of replicates to produce accurate significance levels.

	\section{Numerical Experiments}\label{chap:numexp}

In this section, we provide evidence supporting the utility of our methodology through numerical experiments that mimic observations commonly encountered in practice. Following \cite{granados2022brain} and \cite{ombao2024spectral}, we consider some latent oscillatory processes $Z^{\Omega}_{d}$, $d = 1, 2, \ldots, D$, based on an appropriate second-order autoregressive representation such that the spectral densities of $Z^{\Omega}_{d}$ concentrate on the frequency band $\Omega$. Precisely, we simulate latent theta, alpha and gamma oscillations (denoted by $Z^{\theta}_{d}$, $Z^{\alpha}_{d}$ and $Z^{\gamma}_{d}$, respectively) at sampling rate of 100 Hz with spectral densities having peaks at frequencies 6, 10 and 37.5 Hz, respectively. Then, we establish different magnitudes of spectral group dependence between $\boldsymbol{X}$ and $\boldsymbol{Y}$ by defining the following cases below.

\begin{figure}
	\centerline{
		\includegraphics[width=\textwidth]{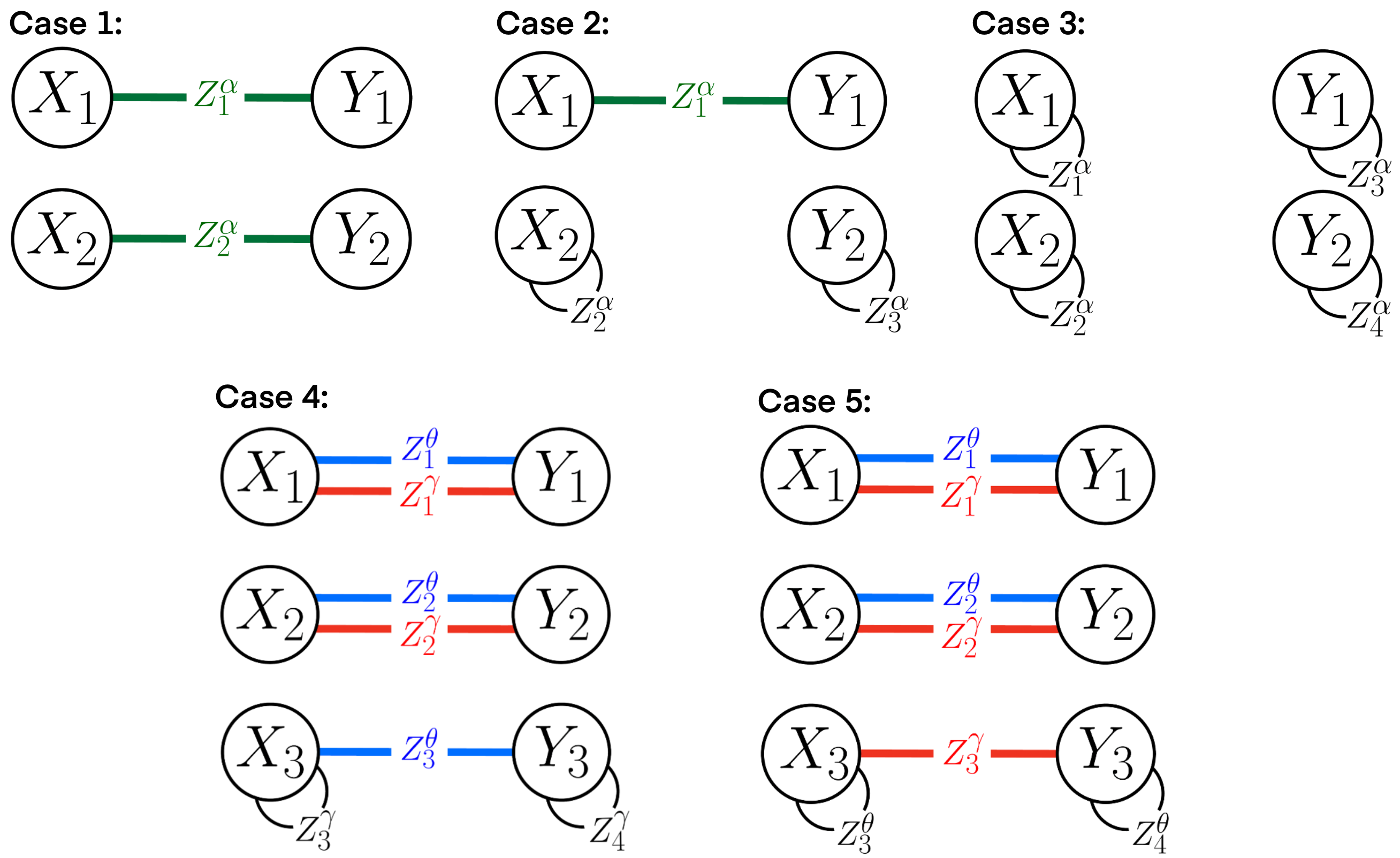}
  }
	\caption{Illustration of the five cases of simulated spectral dependence between two vector time series $\boldsymbol{X}$ and $\boldsymbol{Y}$. Lines connecting two variables represent the shared latent oscillatory component which drives the magnitude of spectral dependence.}
	\label{fig:sim_illus}
\end{figure}

Let $h(\cdot;\cdot)$ be a linear mixing function, i.e., $h(\boldsymbol{Z};\boldsymbol{\phi}) = \boldsymbol{\phi}^\top \boldsymbol{Z}$, where $\boldsymbol{\phi} = (\phi_1,\ldots,\phi_D)^T$ and $\boldsymbol{Z} = (Z_1,\ldots,Z_D)^T$ with $\phi_d > 0$ for all $d$, such that $\sum^D_{d=1} \phi_d = 1$. Also, let $\varepsilon_{d}$ be independent standard normal error processes. For simplicity, we write ``$\{U_1, U_2, \ldots\} \sim h(\boldsymbol{Z};\boldsymbol{\phi})$'' to imply that $U_1, U_2, \ldots,$ are independent realizations from the distribution of $h(\boldsymbol{Z};\boldsymbol{\phi})$. We consider
\begin{enumerate}
    \item[Case 1:] $\{X_{1},Y_{1}\} \sim h((Z^{\alpha}_{1}, \varepsilon_{1})^\top ; \boldsymbol{\phi})$ and $\{X_{2},Y_{2}\} \sim h((Z^{\alpha}_{2}, \varepsilon_{2})^\top; \boldsymbol{\phi})$ with $\boldsymbol{\phi} = (0.75,0.25)^\top$,
    \item[Case 2:] $\{X_{1},Y_{1}\} \sim h((Z^{\alpha}_{1}, \varepsilon_{1})^\top; \boldsymbol{\phi})$, $\{X_{2}\} \sim h((Z^{\alpha}_{2}, \varepsilon_{2})^\top ; \boldsymbol{\phi})$ and $\{Y_{2}\} \sim h((Z^{\alpha}_{3}, \varepsilon_{3})^\top; \boldsymbol{\phi})$ with $\boldsymbol{\phi} = (0.75,0.25)^\top$,
    \item[Case 3:] $\{X_{d}\} \sim h((Z^{\alpha}_{d}, \varepsilon_{d})^\top; \boldsymbol{\phi})$ and $\{Y_{d}\} \sim h((Z^{\alpha}_{d+2}, \varepsilon_{d+2})^\top; \boldsymbol{\phi})$ for $d = 1, 2$, \\with $\boldsymbol{\phi}~=~(0.75,0.25)^\top$,
    \item[Cases 4:] $\{X_{d},Y_{d}\} \sim h((Z^{\theta}_{d}, Z^{\gamma}_{d}, \varepsilon_{d})^\top; \boldsymbol{\phi})$, $d = 1, 2$, $\{X_{3}\} \sim h((Z^{\theta}_{3}, Z^{\gamma}_{3}, \varepsilon_{3})^\top; \boldsymbol{\phi})$ and \\$\{Y_{3}\} \sim h((Z^{\theta}_{3}, Z^{\gamma}_{4}, \varepsilon_{4})^\top; \boldsymbol{\phi})$ with $\boldsymbol{\phi} = (0.375,0.375,0.25)^\top$,
    \item[Cases 5:] $\{X_{d},Y_{d}\} \sim h((Z^{\theta}_{d}, Z^{\gamma}_{d}, \varepsilon_{d})^\top; \boldsymbol{\phi})$, $d = 1, 2$, $\{X_{3}\} \sim h((Z^{\theta}_{3}, Z^{\gamma}_{3}, \varepsilon_{3})^\top; \boldsymbol{\phi})$ and \\$\{Y_{3}\} \sim h((Z^{\theta}_{4}, Z^{\gamma}_{3}, \varepsilon_{4})^\top; \boldsymbol{\phi})$ with $\boldsymbol{\phi} = (0.375,0.375,0.25)^\top$.
\end{enumerate}
\noindent Then, let $\boldsymbol{X} = (X_{1},X_{2},\ldots,X_{p})^\top$ and $\boldsymbol{Y} = (Y_{1},Y_{2},\ldots,Y_{q})^\top$, where $p = q = 2$ for cases 1, 2 and 3, while $p = q = 3$ for cases~4 and 5. Figure~\ref{fig:sim_illus} illustrates, for each case, the simulated spectral dependence between $\boldsymbol{X}$ and $\boldsymbol{Y}$, driven by the shared latent oscillations between their respective components. Here, we set the coefficients $\boldsymbol{\phi}$ such that the signal-to-noise ratio is around 75\%, which is typically observed in real EEG data applications, and for cases~4 and~5, the latent theta and gamma oscillations to have equal weights.

\begin{figure}
	\centerline{
		\includegraphics[width=0.7\textwidth]{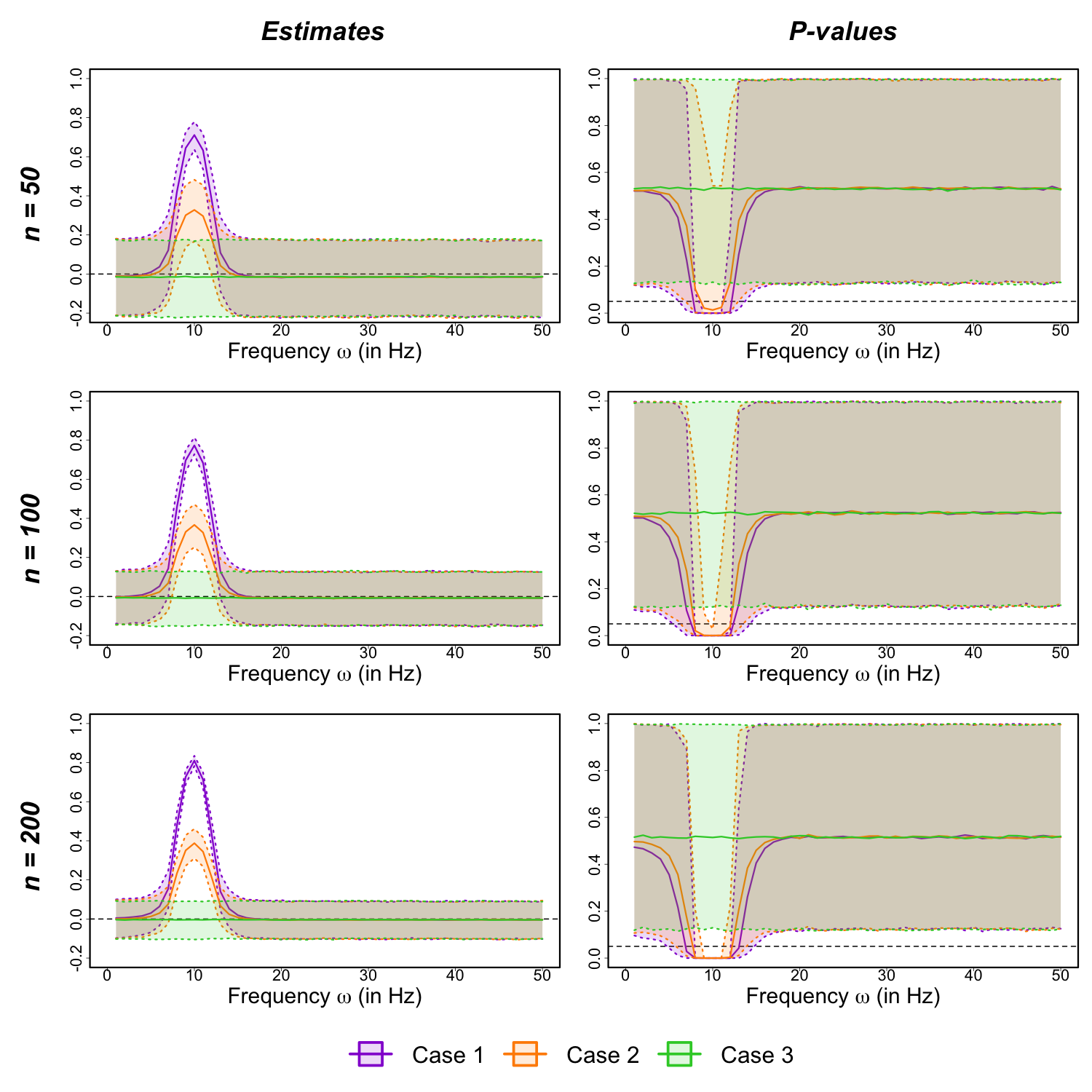}
  }
	\caption{Nonlinear vector coherence estimates (left) and p-values (right) for \textbf{cases 1, 2} and \textbf{3} of spectral dependence over increasing sample sizes (top to bottom). The solid lines correspond to the average of estimates while shaded areas correspond to the middle 95\% quantiles across all 5000 Monte Carlo replicates for each scenario. The black dashed lines indicate the value zero for the estimates (left) and the nominal level 0.05 for p-value comparisons (right).}
	\label{fig:case1}
\end{figure}

\begin{figure}
	\centerline{
		\includegraphics[width=0.7\textwidth]{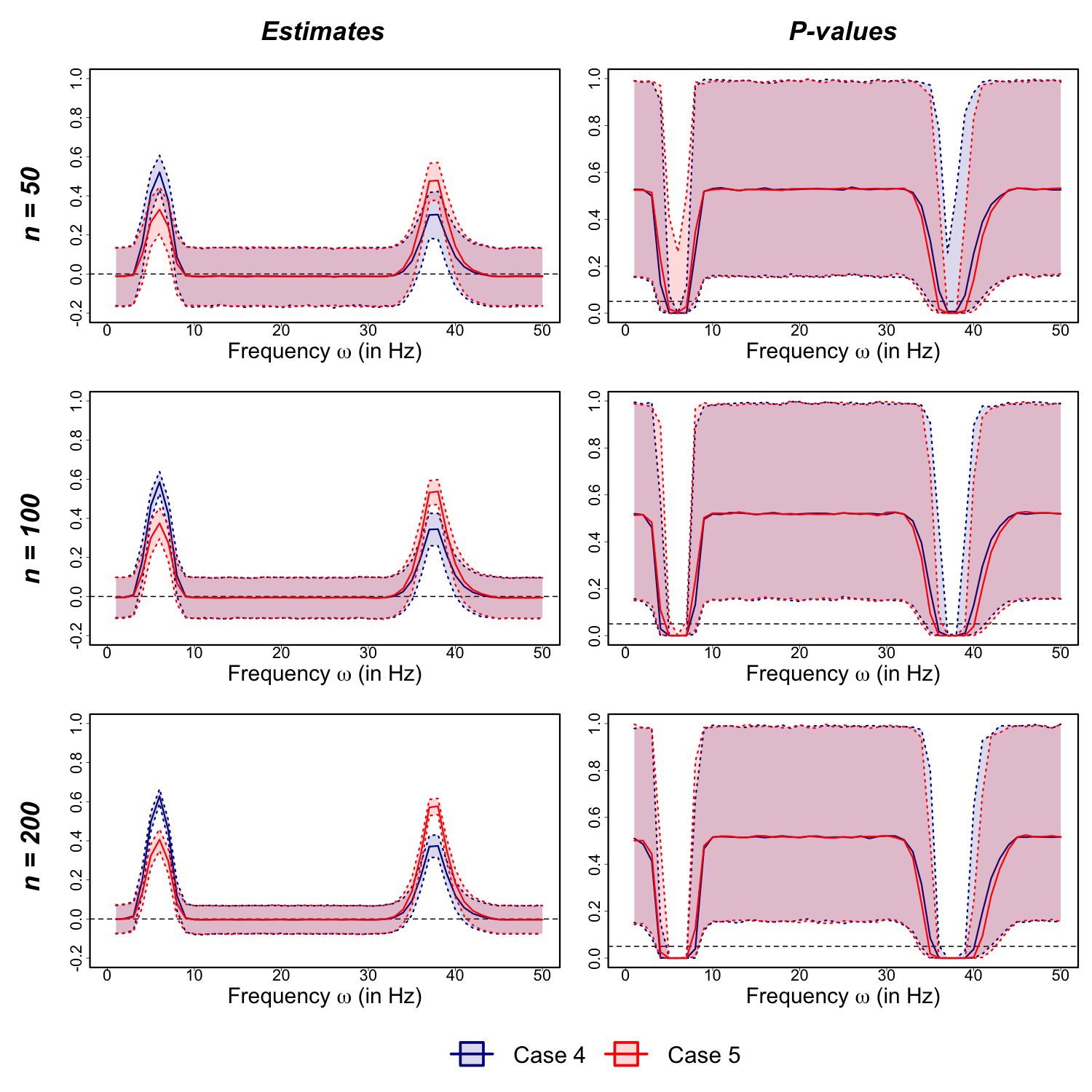}
  }
	\caption{Nonlinear vector coherence estimates (left) and p-values (right) for \textbf{cases 4} and \textbf{5} of spectral dependence over increasing sample sizes (top to bottom). The solid lines correspond to the average of estimates while shaded areas correspond to the middle 95\% quantiles across all 5000 Monte Carlo replicates for each scenario. The black dashed lines indicate the value zero for the estimates (left) and the nominal level 0.05 for p-value comparisons (right).}
	\label{fig:case2}
\end{figure}

Intuitively, the strength of simulated spectral dependence between $\boldsymbol{X}$ and $\boldsymbol{Y}$ is proportional to the number of components that share the same latent oscillations. For example, in the first case, $X_{1}$ and $Y_{1}$ share the same latent alpha-oscillation $Z^{\alpha}_{1}$, while $X_{2}$ and $Y_{2}$ share the same $Z^{\alpha}_{2}$ latent process. Thus, by construction, Cases 1 and 2 exhibit strong and moderate dependence, respectively, in the alpha band while Case 3 represent spectral independence at any frequency. On the other hand, Cases 4 and 5 both reflect dependence at the theta and gamma bands where for the former, the magnitude of dependence is larger in the theta band than in the gamma band, while for the latter, the reverse holds true. For each case, we simulate observations of length $n$ seconds at a sampling rate of 100 Hz, where $n \in \{50,100,200\}$, and employ our methodology (assuming a block length $B = 100$) to estimate the magnitude of spectral dependence and test for spectral independence. Finally, the numerical experiment was repeated for 5000 Monte Carlo replicates to obtain the empirical behavior of the rank-based estimator and our nonparametric test. The results are summarized in Figures~\ref{fig:case1}~and~\ref{fig:case2}, and detailed in Table~\ref{tab:numexp}.

\begin{table}
\caption{Empirical performance of the proposed estimator for the nonlinear vector coherence measure and the nonparametric test of spectral independence. For each simulated case, (i) the standard error of the proposed estimator, and (ii) the probability of rejecting the null hypothesis of spectral independence ($H_0$), averaged over pre-specified sets of frequencies, are computed over 5,000 Monte Carlo replicates based on increasing sample sizes ($n = 50, 100$ and $200$ seconds). Frequency intervals, where spectral dependence is induced, are marked with $(*)$.}
\label{tab:numexp}
\small
\begin{center}
\begin{tabular}{ccccc||ccc}

\multirow{2}{*}{\textbf{Cases}} & \multirow{2}{*}{\textbf{Frequencies (in Hz)}} & \multicolumn{3}{c}{\textbf{Ave. Standard Error}} & \multicolumn{3}{c}{\textbf{Prob. of Rejecting $H_0$}}\\
& & $n=50$ & $n=100$ & $n=200$ &  $n=50$ & $n=100$ & $n=200$ \\
\hline
\hline              
\multirow{2}{*}{Case 1}& $\in (8,12]~(*)$ & 0.0517 & 0.0339 & 0.0225 & 0.9944 & 1.0000 & 1.0000 \\
& $\notin (8,12]$ & 0.0997 & 0.0699 & 0.0491 & 0.0416 & 0.0559 & 0.0770 \\
\hline
\multirow{2}{*}{Case 2}& $\in (8,12]~(*)$ & 0.0842 & 0.0580 & 0.0401 & 0.7650 & 0.9476 & 0.9976 \\
& $\notin (8,12]$ & 0.1001 & 0.0702 & 0.0491 & 0.0228 & 0.0365 & 0.0458 \\
\hline
\multirow{2}{*}{Case 3}& $\in (8,12]$ & 0.0995 & 0.0704 & 0.0496 & 0.0106 & 0.0112 & 0.0119 \\
& $\notin (8,12]$ & 0.1000 & 0.0704 & 0.0493 & 0.0107 & 0.0117 & 0.0113 \\
\hline
\multirow{3}{*}{Case 4}& $\in (4,8]~(*)$ & 0.0586 & 0.0400 & 0.0273 & 0.7768 & 0.8393 & 0.9390 \\
& $\in (35,40]~(*)$ & 0.0686 & 0.0471 & 0.0328 & 0.6202 & 0.8456 & 0.9532 \\
& $\notin (4,8] \cup (35,40]$ & 0.0764 & 0.0534 & 0.0370 & 0.0172 & 0.0335 & 0.0486 \\
\hline
\multirow{3}{*}{Case 5}& $\in (4,8]~(*)$ & 0.0670 & 0.0466 & 0.0321 & 0.6938 & 0.7848 & 0.8401 \\
& $\in (35,40]~(*)$ & 0.0609 & 0.0418 & 0.0287 & 0.8337 & 0.9545 & 0.9980 \\
& $\notin (4,8] \cup (35,40]$ & 0.0764 & 0.0533 & 0.0369 & 0.0151 & 0.0338 & 0.0647 \\
\hline
\hline \\~\\
\end{tabular}
\end{center}

\end{table}

Figure~\ref{fig:case1} demonstrates that the proposed estimator successfully captures the magnitudes of spectral dependence across all cases. Estimates from the first two cases attain their peak values at frequency 10 Hz with larger magnitudes for Case 1 and have close-to-zero values for frequencies outside the alpha band. Moreover, estimates from Case 3 are effectively zero, which indicates independence of $\boldsymbol{X}$ and $\boldsymbol{Y}$ at all frequencies. Furthermore, there is a visible reduction in the uncertainty of the estimates given more observations, which roughly decreases at the same rate as the square root of the sample size, as typically expected (see Table~\ref{tab:numexp}). That is, as the sample size increases, we observe the middle 95\% empirical quantiles of the estimates to shrink towards the mean estimates (see Figure~\ref{fig:case1}, left panels). This hints at the consistency of our estimation procedure. Additionally, the corresponding p-values based on the proposed nonparametric test are generally smaller than the standard nominal level of 0.05 for significant cases, while remaining large (centered around 0.5) for independent cases (see Figure~\ref{fig:case1}, right panels). Performance of our nonparametric test also improves with an increase in statistical power for detecting significant dependence as the sample size increases, while maintaining approximately correct sizes under spectral independence, as shown in Table~\ref{tab:numexp}.

We observe a similar behavior in Figure~\ref{fig:case2}. In particular, our NVC estimator achieves larger magnitudes for stronger spectral dependence at the correct frequencies. For example, estimates attain their peaks at frequencies 6 and 37.5 Hz, as simulated in both cases, and the estimated magnitudes are higher in the theta band for Case 4, while it is higher in the gamma band for Case 5. Furthermore, close-to-zero estimates are observed for frequencies between the two simulated frequency bands, which confirms simulated spectral independence. In addition, associated p-values are smaller than the specified 0.05 nominal level in the alpha band for cases 1--3 and in the theta and gamma bands for cases 4 and 5, and are larger for frequencies where there is spectral independence for all cases. Lastly, the effect of having larger sample sizes in reducing estimate uncertainties, i.e., the shrinking of middle 95\% empirical quantiles of the estimates towards the mean estimates (see Figure~\ref{fig:case2}, left panels), and the statistical power improvement while maintaining approximately correct sizes (see Table~\ref{tab:numexp}), are also evident. Hence, the proposed inference approach for the new nonlinear vector coherence measure offers a novel yet easy-to-implement statistical framework for exploring functional connectivity between groups of signals originating from different brain regions.

	\section{NVC Analysis of Alzheimer's Disease and Frontotemporal Dementia Resting-state EEG Data}\label{chap:eeganalysis}

\subsection{Exploring Region-to-region Functional Connectivity}

In this paper, one of our primary objectives is to understand the functional connectivity alterations between different brain regions (recall Figure~\ref{fig:eeg_rois}) that are associated with AD and FTD. To address this goal, we employ three methodologies for quantifying dependence, namely, the $T$ measure defined in Equation~(\ref{eq:T_AnF}), the PBC measure defined in Equation~(\ref{eq:PBC}) and our novel NVC measure defined in Equation~(\ref{eq:nlcoh_vec}). The $T$ measure, which we refer to as the non-spectral vector connectivity (nSCV) approach from here on, allows for investigating region-to-region functional connectivity. However, given that it is non-spectral, it has the limitation that it is not able to attribute to which frequency oscillation the captured nonlinear dependence manifests. On the contrary, PBC measures dependence in the frequency domain but is formulated only for a pair of univariate time series (i.e., signals from individual channels) rather than between vector time series (i.e., signals from groups of channels). Thus, it further requires some form of aggregation across computed pairwise quantities to capture dependence between brain regions.

Our NVC measure is tailored to address the limitations of the nSVC and PBC approaches, i.e., it enables for exploring nonlinear spectral dependence between ROIs. More specifically, as NVC is naturally formulated as a measure of predictability for vector time series in the frequency domain, it directly captures general (beyond linear) spectral dependence between a pair of groups of channels, which does not require arbitrary aggregation of pairwise dependencies. Hence, it enjoys the desirable properties of the two approaches combined. By comparing findings derived by NVC to the results based on nSVC and PBC, another objective is to highlight the functional connectivity features that our novel NVC approach quantifies more appropriately, and the connections that the two approaches fail to characterize due to their respective limitations. Such comparisons help improve our understanding of the neurological dysfunctions associated with AD and FTD, thereby providing evidence of the practical utility of the proposed methodology in neurology.

For each individual, we estimate the three quantities using all available EEG data excluding the first five seconds of recordings for stability purposes. This is to avoid inclusion of brain activity that may not be at a ``resting-state", which typically occurs during the beginning of the experiment. Moreover, when computing the nSVC and NVC measures, we use their respective symmetric permutation-invariant versions as functional connectivity between ROIs is non-directional. We compute the NVC measure with $B=100$ blocks (i.e., obtaining periodograms over one-second time segments) to ensure that a sufficient number of observations is available for estimation, while maintaining an adequate number of individual frequencies being represented. In calculating the PBC measure for each of the five canonical frequency bands (i.e., the delta, theta, alpha, beta and gamma bands), we employ the estimation procedure based on the maximum squared lagged cross-correlation between band-specific filtered signals, as described in \cite{ombao2024spectral}, with a maximum lag of 50. Lastly, because the NVC measure is defined for individual frequencies, we summarize the NVC values over all frequencies in each of the five frequency bands to facilitate comparable interpretations with the PBC approach.

\begin{figure}
	\centerline{
		\includegraphics[width=0.55\textwidth]{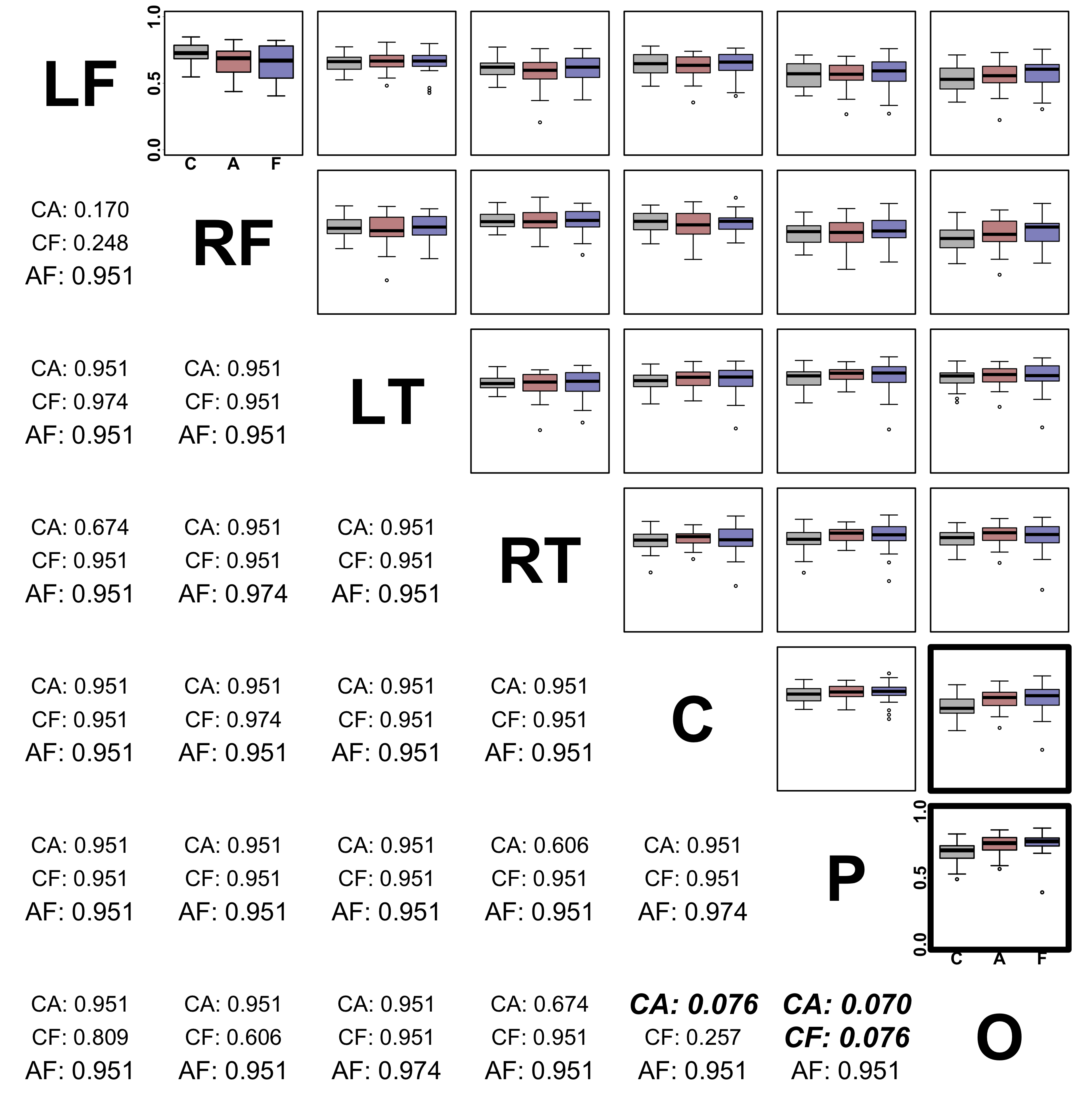}
  }
	\caption{Distribution of estimated functional connectivity (based on \textbf{nSVC}) across subjects in the CN, AD and FTD groups. The plots in the upper triangle display the boxplots of dependence estimates from the subjects in the CN (left), AD (middle) and FTD (right) groups for each pair of ROIs. The numbers in the lower triangle indicate the corrected p-values for the comparison pairs (CA: CN vs. AD; CF: CN vs. FTD; and AF: AD vs. FTD). P-values less than 0.05 and their corresponding boxplots are highlighted in bold.}
	\label{fig:eeg_results_AF}
\end{figure}

\begin{figure}
    \footnotesize
    \textbf{~~~~~~~(a) Delta}~~~~~~~~~~~~~~~~~~~~~~~~~~~~~~~~~~~~~~~~~~~~~~~~~~~~~~\textbf{(b) Theta}
    
    \centerline{
    \includegraphics[width=0.4\textwidth]{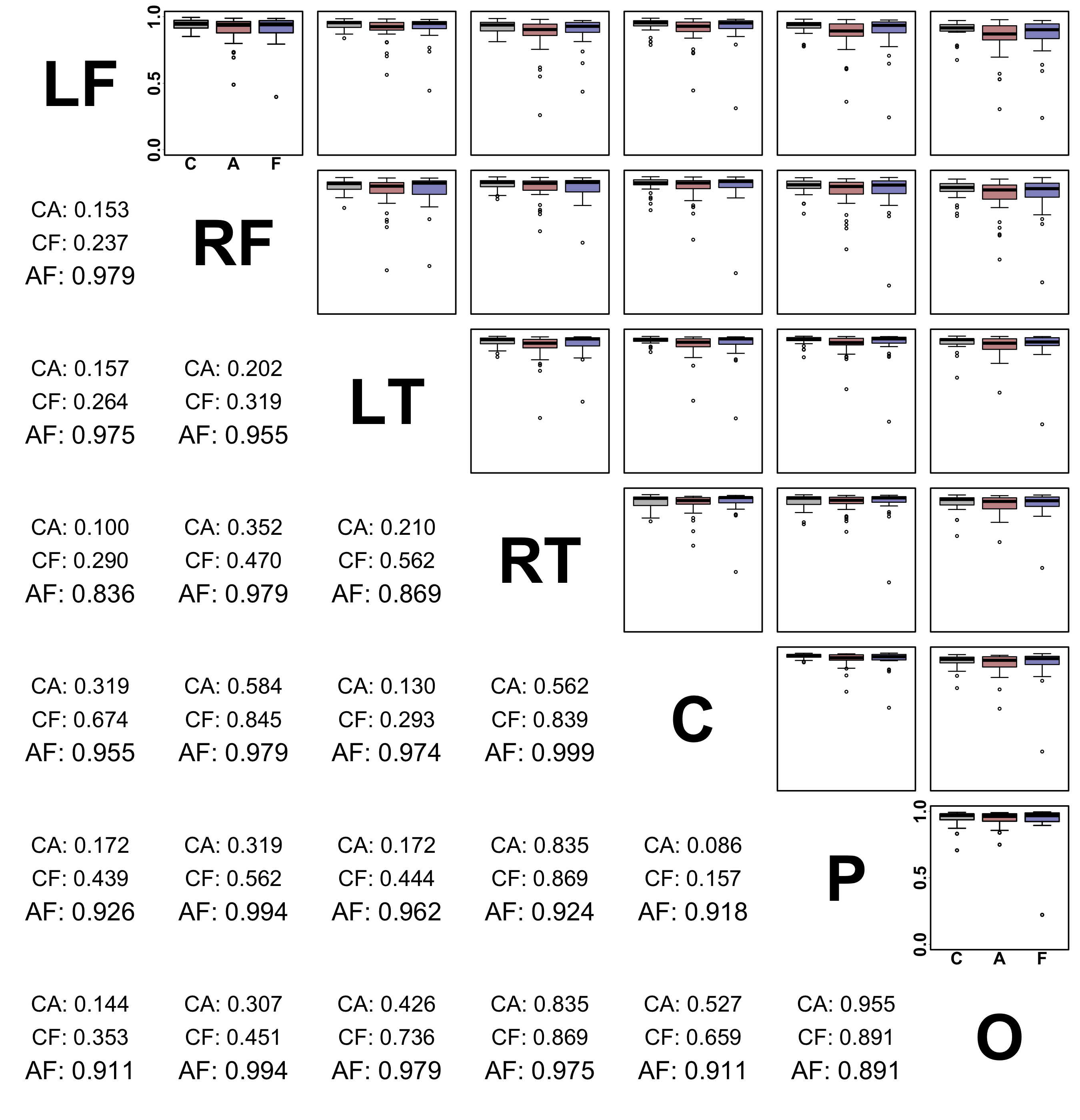} \includegraphics[width=0.4\textwidth]{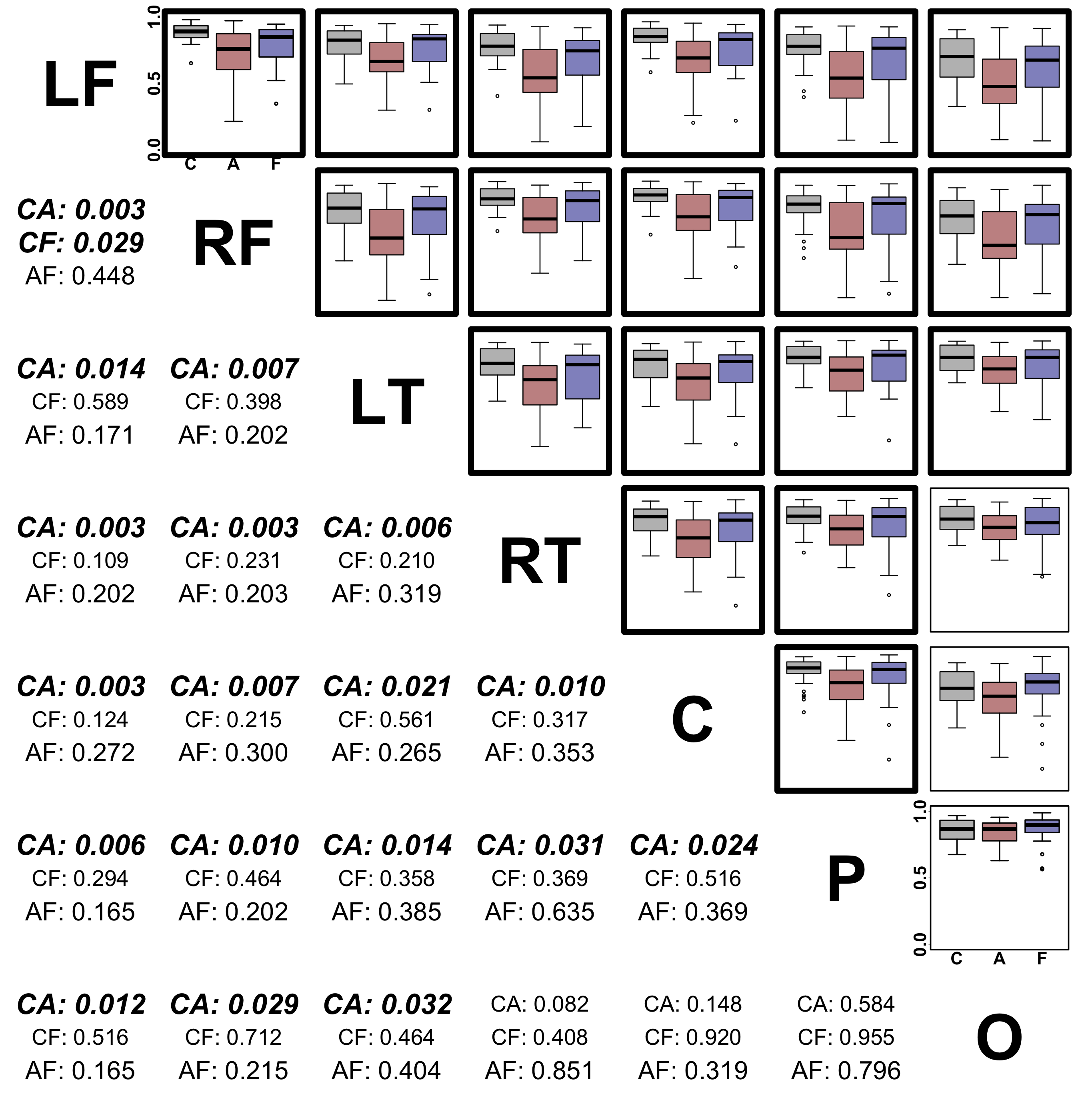}\\~\\
    }

    \textbf{~~~~~~~(c) Alpha}~~~~~~~~~~~~~~~~~~~~~~~~~~~~~~~~~~~~~~~~~~~~~~~~~~~~~\textbf{(d) Beta}
    
    \centerline{
    \includegraphics[width=0.4\textwidth]{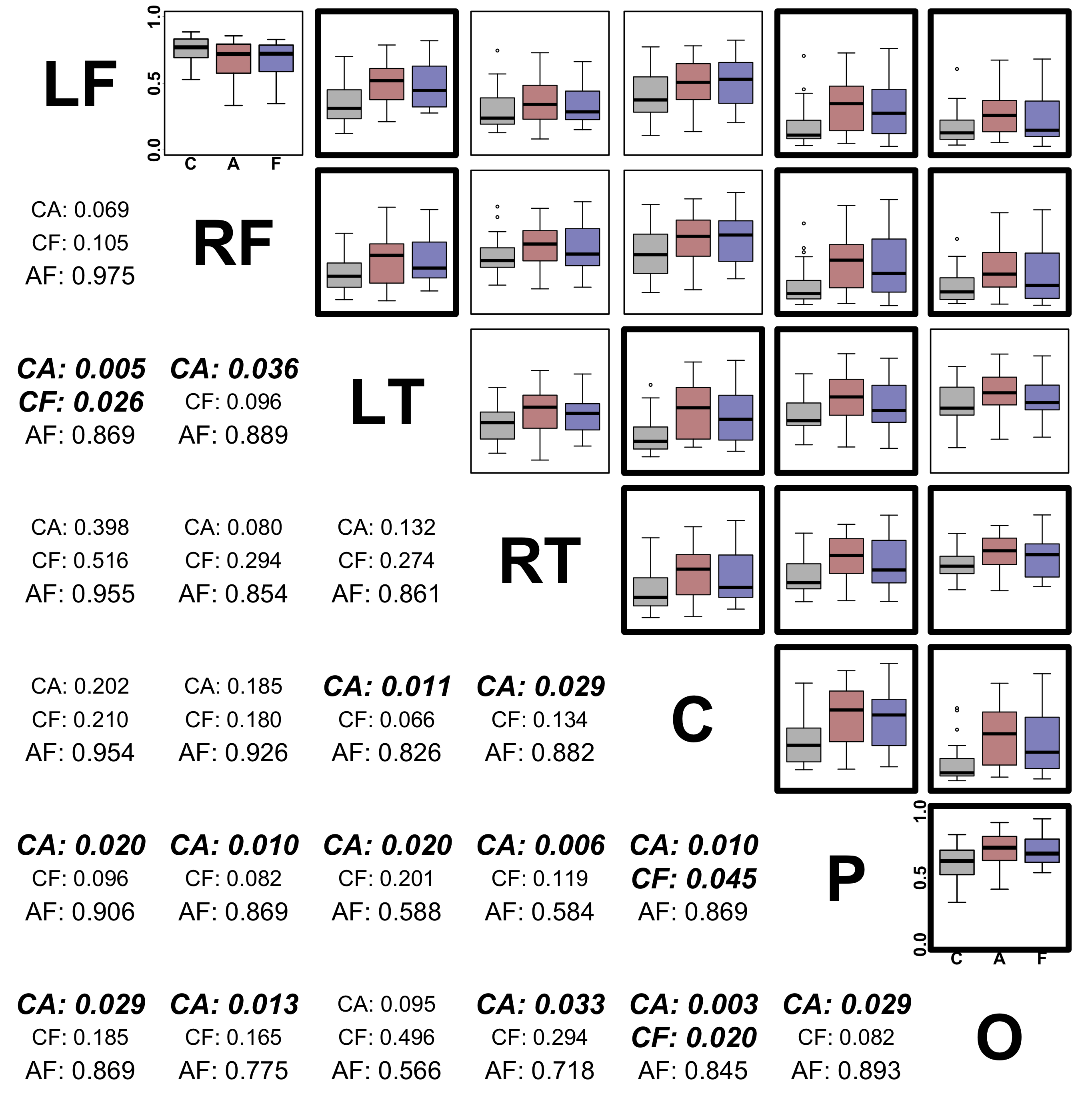} \includegraphics[width=0.4\textwidth]{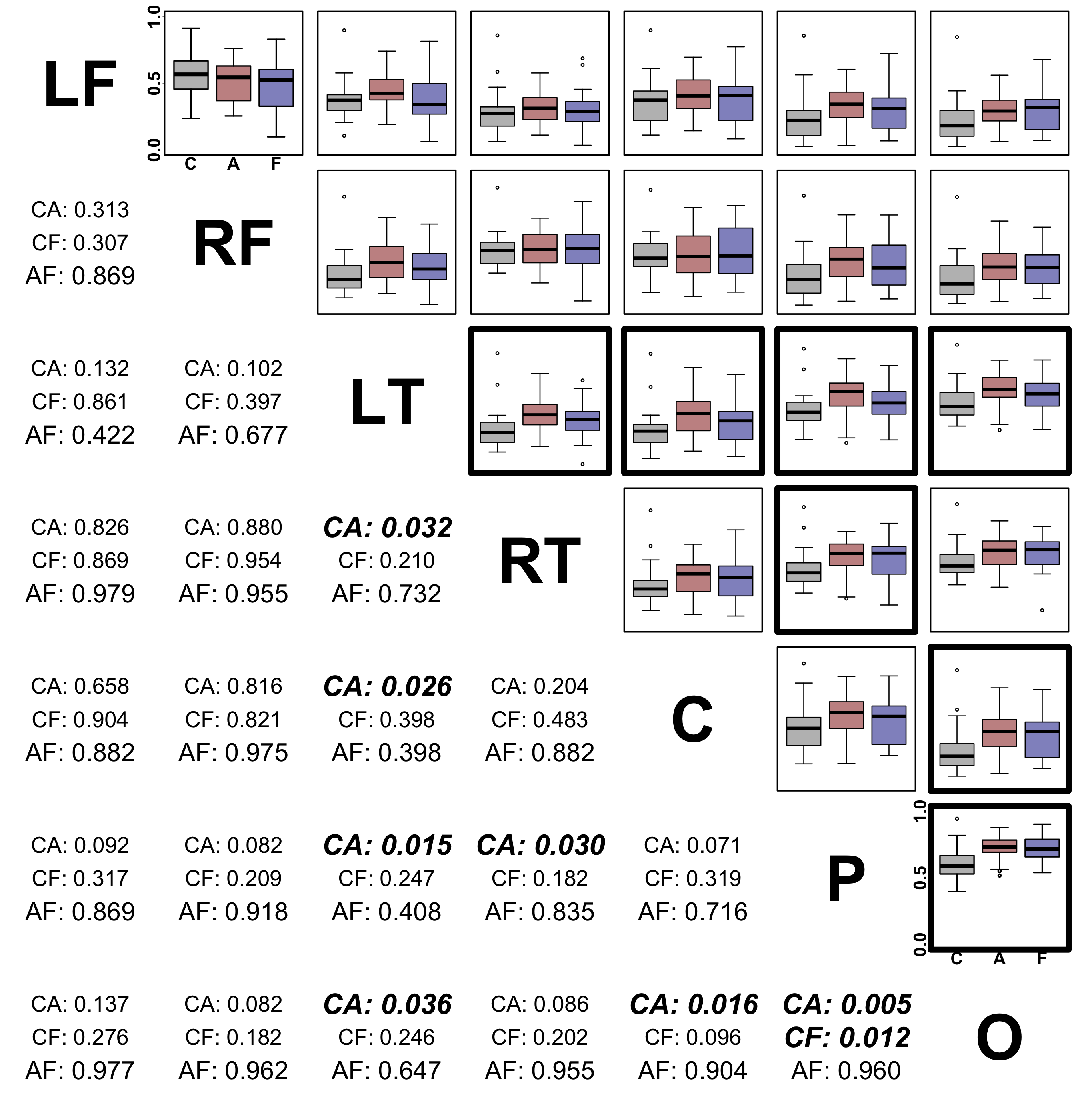} \\~\\
    }

    \textbf{~~~~~~~~~~~~~~~~~~~~~~~~~~~~~~~~~~~~~~~~~~~(e) Gamma}
    
    \centerline{
    \includegraphics[width=0.4\textwidth]{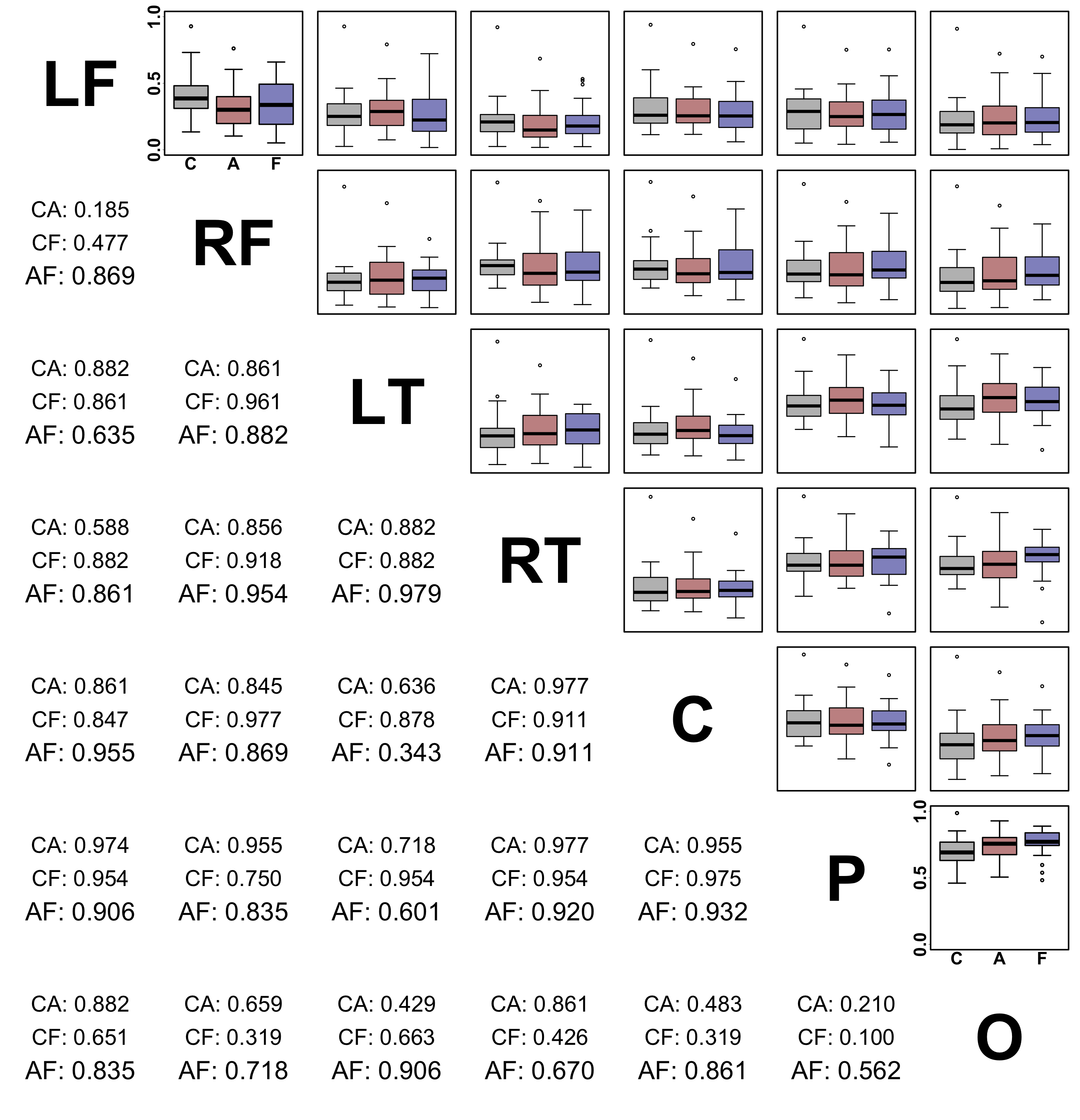}
    }
	\caption{Distribution of estimated functional connectivity (based on \textbf{PBC}) for the delta (a), theta (b), alpha (c), beta (d), and gamma (e) bands across subjects in the CN, AD and FTD groups. The plots in the upper triangle display the boxplots of dependence estimates from the subjects in the CN (left), AD (middle) and FTD (right) groups for each pair of ROIs. The numbers in the lower triangle indicates the corrected p-values for the comparison pairs (CA: CN vs. AD; CF: CN vs. FTD; and AF: AD vs. FTD). P-values less than 0.05 and their corresponding boxplots are highlighted in bold.}
	\label{fig:eeg_results_PBC}
\end{figure}

\begin{figure}
    \footnotesize
    \textbf{~~~~~~~(a) Delta}~~~~~~~~~~~~~~~~~~~~~~~~~~~~~~~~~~~~~~~~~~~~~~~~~~~~~~\textbf{(b) Theta}
    
    \centerline{
    \includegraphics[width=0.4\textwidth]{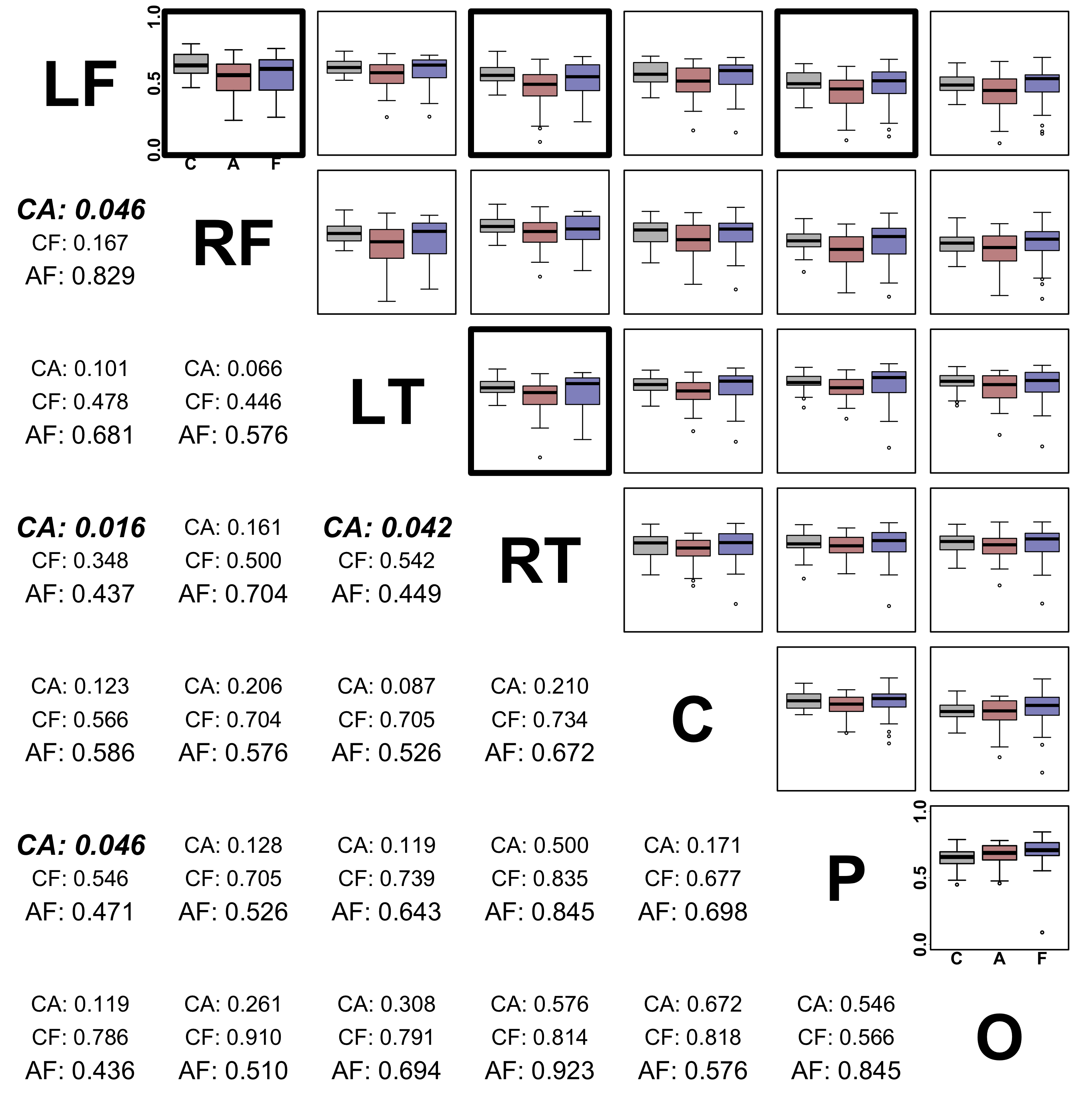} \includegraphics[width=0.4\textwidth]{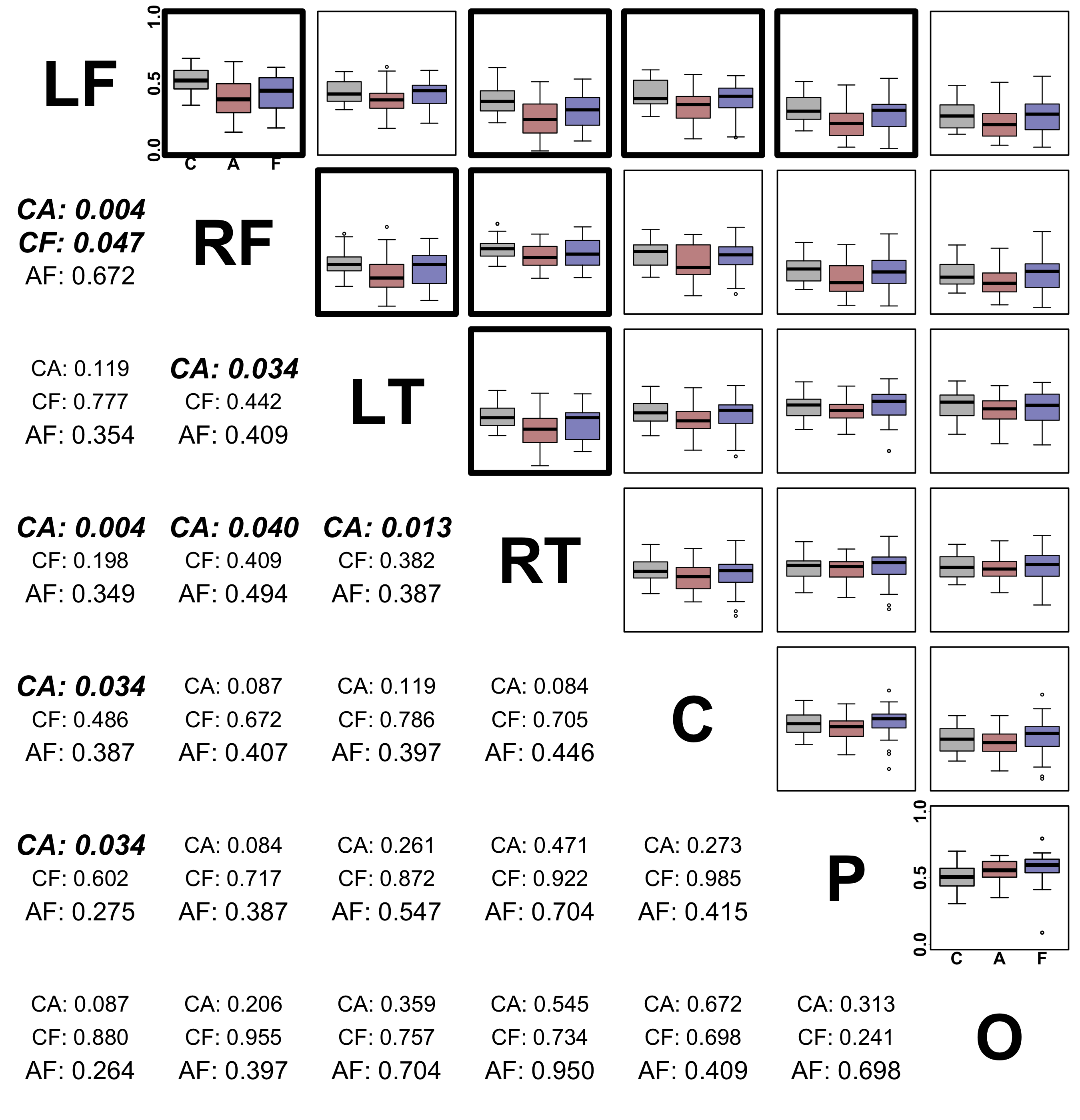}\\~\\
    }

    \textbf{~~~~~~~(c) Alpha}~~~~~~~~~~~~~~~~~~~~~~~~~~~~~~~~~~~~~~~~~~~~~~~~~~~~~\textbf{(d) Beta}
    
    \centerline{
    \includegraphics[width=0.4\textwidth]{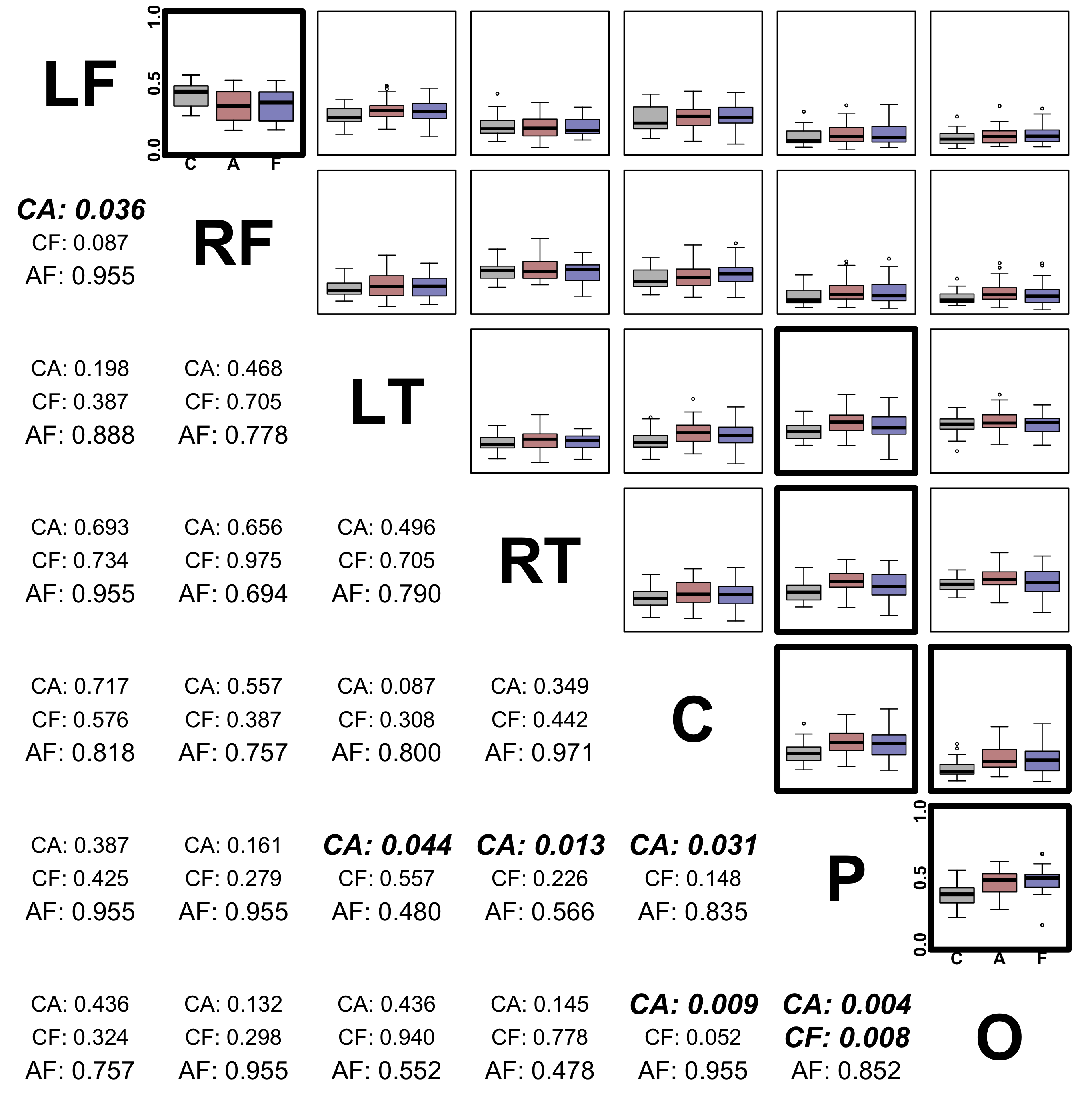} \includegraphics[width=0.4\textwidth]{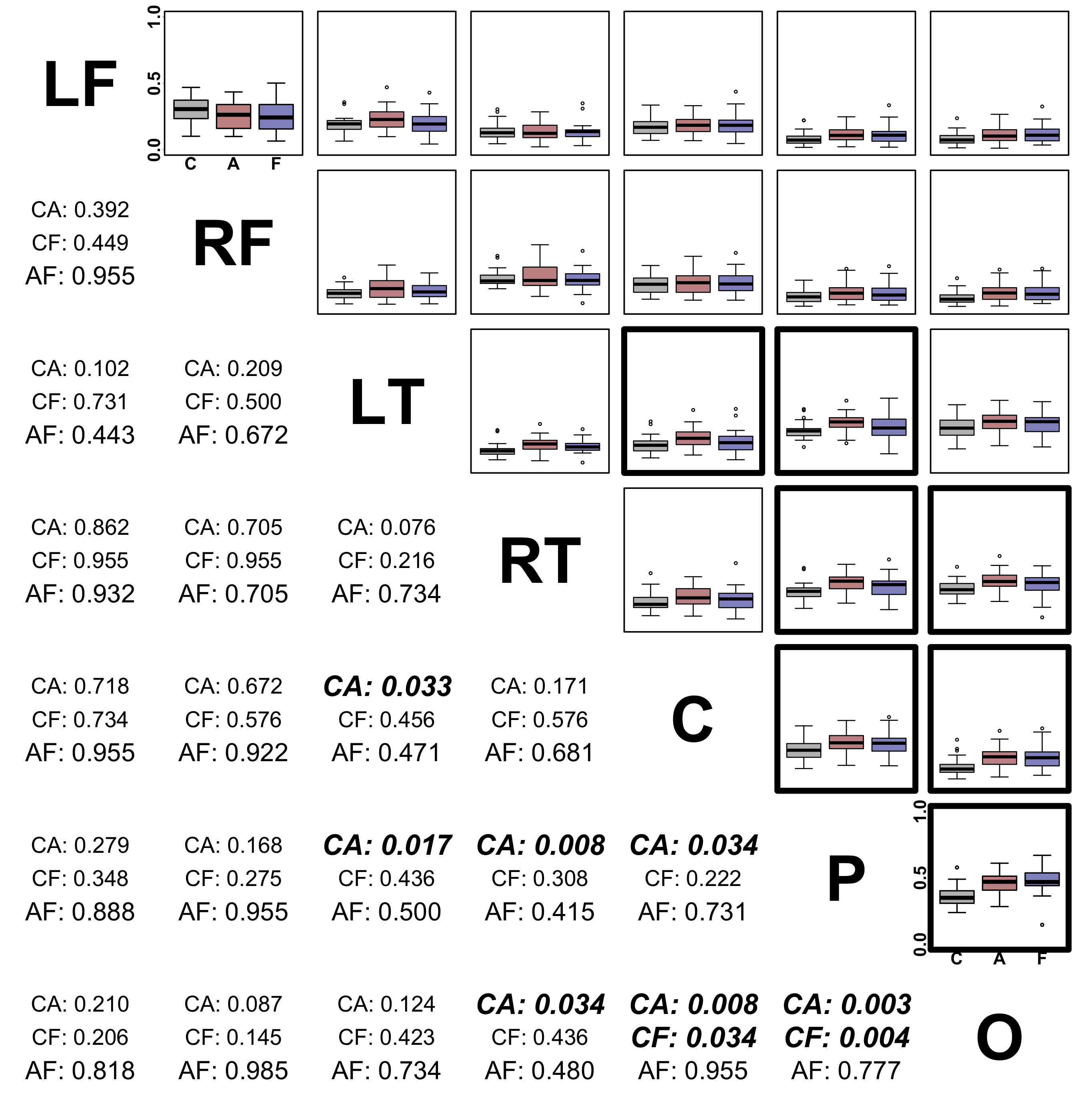} \\~\\
    }

    \textbf{~~~~~~~~~~~~~~~~~~~~~~~~~~~~~~~~~~~~~~~~~~~(e) Gamma}
    
    \centerline{
    \includegraphics[width=0.4\textwidth]{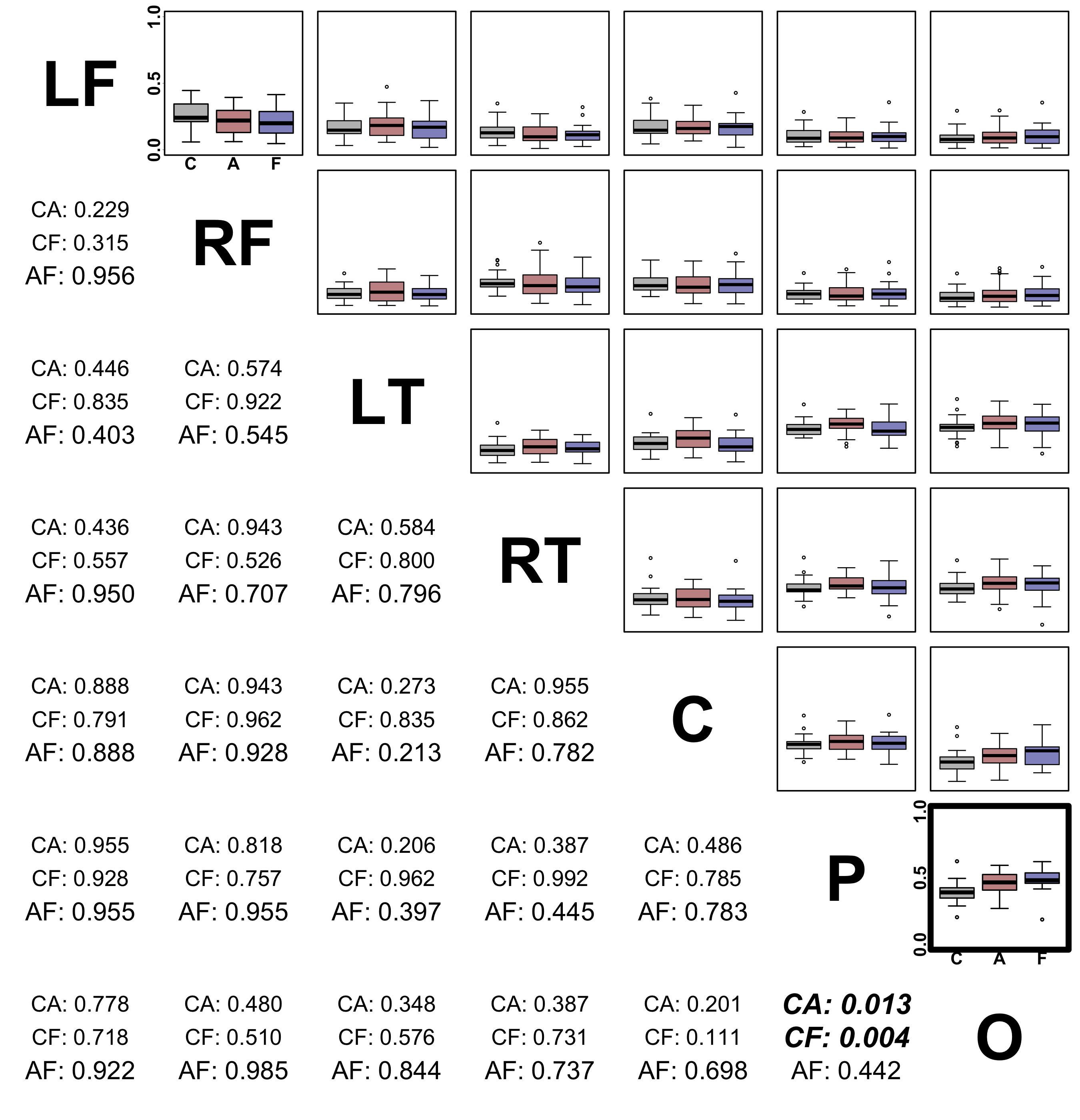}
    }
	\caption{Distribution of estimated functional connectivity (based on \textbf{NVC}) for the delta (a), theta (b), alpha (c), beta (d), and gamma (e) bands across subjects in the CN, AD and FTD groups. The plots in the upper triangle display the boxplots of dependence estimates from the subjects in the CN (left), AD (middle) and FTD (right) groups for each pair of ROIs. The numbers in the lower triangle indicates the corrected p-values for the comparison pairs (CA: CN vs. AD; CF: CN vs. FTD; and AF: AD vs. FTD). P-values less than 0.05 and their corresponding boxplots are highlighted in bold.}
	\label{fig:eeg_results_NVC}
\end{figure}

We then compare the distribution of these dependence measures for all ROI pairs across the three subject groups (see Figures~\ref{fig:eeg_results_AF}--\ref{fig:eeg_results_NVC}). Specifically, we test for significant differences in functional connectivity among the groups, which includes three comparison pairs (namely, CN vs. AD, CN vs. FTD and AD vs. FTD), using standard permutation tests. To control for the inflation of Type I errors due to multiple comparisons, we adjust the p-values using the Benjamini--Hochberg (BH) procedure \citep{benjamini1995controlling}. This results in 63 corrected p-values (21 ROI pairs times 3 comparison pairs) for the nSVC approach and 315 corrected p-values (21 ROI pairs times 5 frequency bands times 3 comparison pairs) for the PBC and NVC approaches. Comparisons resulting in corrected p-values less than 0.05 are considered significant differences in functional connectivity between the associated groups.

In Figure~\ref{fig:eeg_results_AF}, we see that the dependence, as captured by nSVC, between the parietal and occipital ($P$--$O$) regions among the healthy controls is significantly lower than among the subjects in the AD and FTD groups. Moreover, the functional connections within the central and occipital ($C$--$O$) regions are only significantly different between the CN and AD groups. Despite such identified differences, the practical relevance of these findings is limited, as nSVC is not able to detect more subtle brain connectivity alterations. Furthermore, the magnitude of dependence derived from the nSVC approach has limited interpretation as it is not tied to any neural oscillations that have established cognitive functions.

By contrast, Figures~\ref{fig:eeg_results_PBC}~and~\ref{fig:eeg_results_NVC} illustrate the distribution of region-to-region functional connectivity by subject group across the five frequency bands, as quantified by the aggregated PBC and our proposed NVC measure. One striking dissimilarity is that the PBC approach does not detect any significant differences in the delta and gamma bands (see Figure~\ref{fig:eeg_results_PBC}, panels (a) and (e)). In addition, the PBC values exhibit high variability across subjects, especially within the AD and FTD groups. Nonetheless, the PBC approach suggests 39 differentiating dependence features between the CN and AD groups, and 5 dependence features that separate the CN and FTD groups. According to these results, PBC indicates that subjects with AD demonstrate significantly lower magnitude of functional connectivity at the theta band, as compared to healthy controls, in the entire brain network (with the exception of some connections in the $RT$--$O$, $C$--$O$ and $P$--$O$ regions). By contrast, the majority of the region-to-region dependencies at the alpha band (i.e., 14 out of 21) are significantly weaker, whereas several at the beta band (i.e., 7 out of 21) are significantly stronger among healthy subjects compared to subjects in the AD group. Comparing now the CN and FTD groups, PBC values between the $LF$ and $RF$ regions in the theta band are significantly larger among the healthy controls while the reverse is true for the $LF$--$LT$, $C$--$P$, and $C$--$O$ regions at the alpha band, and the $P$--$O$ region at the beta band.

Our NVC approach, on the other hand, reveals a richer (and quite different) set of findings. Precisely, the NVC measure captures 25 distinguishing dependence features between the CN and AD groups, and five significantly different functional connectivity patterns among healthy subjects and subjects with FTD, which manifest throughout the five canonical frequency bands (see Figure~\ref{fig:eeg_results_NVC}). Between the healthy controls and subjects in the AD group, we observe significantly different NVC values within the left and right lateral brain regions at low frequencies (which includes the delta and theta bands). To be more precise, subjects with AD exhibit weaker functional connectivity at low frequencies, between the left and right frontal regions, which is evident even at the alpha band. Dependence between the two temporal regions (i.e., $LT$ and $RT$) and functional connections of LF and RF regions with other brain regions, as measured by NVC, are also significantly weaker at the delta and theta bands among subjects with AD compared to healthy subjects. Another feature that separates the CN group from the AD group is that several dependencies within the posterior regions are significantly weaker among the healthy controls at high frequencies (including the alpha, beta and gamma bands). Lastly, our NVC measure also demonstrates differentiating features between the CN and FTD groups, i.e., healthy subjects exhibit stronger functional dependence between the two frontal regions at the theta band while subjects with FTD have stronger connections in the $C$--$O$ region at the beta band, and in the $P$--$O$ region at the alpha, beta and gamma bands.

As for finding distinguishing functional connectivity patterns between the AD and the FTD groups, none of the three methodologies yield significant differences. This may be due to the two degenerative conditions having very similar neurological dysfunctions, which only manifest at later stages of their respective progressions \citep{musa2020alzheimer}. Nonetheless, other works that investigated functional connectivity using the same dataset arrived at the same conclusion (see, e.g., \citealp{wu2024changes} and \citealp{zheng2025time}).

\subsection{Evaluating Extracted Functional Connectivity Features through Classification}

Even though the PBC and NVC approaches reveal region-to-region connectivity patterns that differentiate healthy subjects from subjects in the AD and FTD groups, the magnitude and type of dependence quantified by the two methodologies are fundamentally different. Unlike NVC, which is naturally formulated to measure nonlinear spectral dependence between vector time series, we emphasize that the aggregated PBC approach has two disadvantages. First, coherence is a linear measure of dependence, and hence, it may not completely quantify functional connections in complex systems such as the human brain. Moreover, since the choice of aggregation method to summarize dependence within two ROIs is arbitrary, this adds another layer of uncertainty in the magnitude of functional connections derived from this approach.

\begin{table}
\small
\caption{Top significant differentiating dependence features between the CN and AD groups and between the CN and FTD groups, as extracted by the PBC and NVC measures. Ranks are based on the lowest p-values. Features with larger magnitude among subjects in the CN group compared to the disease groups are indicated with $(+)$.}
\label{tab:topdep}
\begin{center}
\begin{tabular}{cccc||ccc}

\multirow{2}{*}{\textbf{Task}} & \multicolumn{3}{c}{\textbf{Nonlinear Vector Coherence}} & \multicolumn{3}{c}{\textbf{Pairwise Band Coherence}}\\
& \textbf{Rank} & \textbf{Feature} & \textbf{Freq. Band}  & \textbf{Rank} & \textbf{Feature} & \textbf{Freq. Band} \\

\hline
\hline
\multirow{10}{*}{\textbf{CN vs. AD}} & 1 & $P$ -- $O$ & $\beta$ & 1 & $~~~~~~LF$ -- $RT~(+)$ & $\theta$ \\
 & 2 & $~~~~~~LF$ -- $RT~(+)$ & $\theta$ & 2 & $C$ -- $O$ & $\alpha$ \\
 & 3 & $~~~~~~LF$ -- $RF~(+)$ & $\theta$ & 3 & $~~~~~~LF$ -- $RF~(+)$ & $\theta$ \\
 & 4 & $P$ -- $O$ & $\alpha$ & 4 & $~~~~~~LF$ -- $C~~~(+)$ & $\theta$ \\
 & 5 & $C$ -- $O$ & $\beta$ & 5 & $~~~~~~RF$ -- $RT~(+)$ & $\theta$ \\
\cline{2-7}\\
 & 6 & $RT$ -- $P~~$ & $\beta$ & 6 & $P$ -- $O$ & $\beta$ \\
 & 7 & $C$ -- $O$ & $\alpha$ & 7 & $LF$ -- $LT$ & $\alpha$ \\
 & 8 & $P$ -- $O$ & $\gamma$ & 8 & $~~~~~~LF$ -- $P~~~(+)$ & $\theta$ \\
 & 9 & $~~~~~~LT$ -- $RT~(+)$ & $\theta$ & 9 & $RT$ -- $P~~$ & $\alpha$ \\
 & 10 & $RT$ -- $P~~$ & $\alpha$ & 10 & $~~~~~~LT$ -- $RT~(+)$ & $\theta$ \\
\hline
\hline
\multirow{5}{*}{\textbf{CN vs. FTD}} & 1 & $P$ -- $O$ & $\beta$ & 1 & $P$ -- $O$ & $\beta$ \\
 & 2 & $P$ -- $O$ & $\gamma$ & 2 & $C$ -- $O$ & $\alpha$ \\
 & 3 & $P$ -- $O$ & $\alpha$ & 3 & $LF$ -- $LT$ & $\alpha$ \\
 & 4 & $C$ -- $O$ & $\beta$ & 4 & $~~~~~~LF$ -- $RF~(+)$ & $\theta$ \\
 & 5 & $~~~~~~LF$ -- $RF~(+)$ & $\theta$ & 5 & $C$ -- $P$ & $\alpha$ \\
\hline
\hline
\end{tabular}
\end{center}

\end{table}

This is evident with respect to the most significant dependence features (i.e., the PBC and NVC values that result in the lowest p-values for the group comparisons) that differentiate the CN group from the AD and FTD groups (see Table~\ref{tab:topdep}). Some of the top dependence features captured by both approaches include the functional connectivity within the $LF$--$RF$, $LF$--$RT$ and $LT$--$RT$ regions at the theta band, $C$--$O$ and $RT$--$P$ regions at the alpha band, and $P$--$O$ region at the beta band. However, the most distinguishing NVC features between the CN and AD groups highlight the functional connectivity within the posterior region (i.e., $RT$, $C$, $P$ and $O$ regions) at high frequencies (i.e., alpha, beta and gamma bands). As for the PBC features, functional connections of the two frontal regions (i.e., $LF$ and $RF$) with other brain regions at the theta band are the most significant. In addition, when differentiating between the CN and FTD groups, the NVC measure extracts three out of five dissimilar dependence features as the PBC approach. Clinical interpretations of these functional connectivity alterations in relation to AD and FTD are further discussed in Section~\ref{subchap:relevance}.

These contrasts lead to different characterizations of the neurological dysfunctions associated with AD and FTD. Thus, to understand which approach produces features that more appropriately discriminate healthy subjects from subjects with AD or with FTD, our solution is to train a classifier (i.e., dicriminating between CN vs. AD and CN vs. FTD subjects) that uses the top dependence features derived by the PBC and NVC measures. Specifically, we employ the support vector machine (SVM) algorithm to find the optimal hyperplane in a high-dimensional space of feature inputs that distinctly discriminate the healthy controls from each of the two disease groups. For a comprehensive review and recent developments on SVMs, see \cite{chauhan2019problem}.

In addition to the top dependence features presented in Table~\ref{tab:topdep}, we also extract complementary marginal features from all ROIs, which may help differentiate the CN group from the AD and FTD groups. To be more precise, we obtain the average relative band power (RBP), i.e., the relative contribution of a frequency band to the total variance of a time series (see, e.g., \citealp{redondo2026statistics}), of the five frequency bands for each subject. Then, we compare the distribution of RBPs among the three subject groups via standard permutation tests, and all resulting p-values are corrected for multiple comparisons via the BH procedure, and are tested at the 0.05 significance level. Details on the exact implementation for RBP (including summaries for each frequency band), and the uncorrected and corrected p-values are reported in the Supplementary Material.

\begin{table}
\small
\caption{Top significant differentiating marginal features between the CN and AD groups and between the CN and FTD groups, as extracted by the average RBP measure. Ranks are based on the lowest p-values. Features with larger magnitude among subjects in the CN group compared to the disease groups are indicated with $(+)$.}
\label{tab:topmar}
\begin{center}
\begin{tabular}{ccc||ccc}

\multicolumn{3}{c}{\textbf{CN vs. AD}} & \multicolumn{3}{c}{\textbf{CN vs. FTD}}\\
\cline{1-3} \cline{4-6}
\textbf{Rank} & \textbf{Feature} & \textbf{Freq. Band}  & \textbf{Rank} & \textbf{Feature} & \textbf{Freq. Band} \\
\hline
\hline              
1 & $~~~~~~~~O~~~(+)$ & $\alpha$ & 1 & $~~~~~~LT~~(+)$ & $\alpha$ \\
2 & $O$ & $\delta$ & 2 & $~~~~~~~O~~~(+)$ & $\alpha$ \\
3 & $~~~~~~LT~~(+)$ & $\alpha$ & 3 & $LF$ & $\gamma$ \\
4 & $~~~~~~RT~~(+)$ & $\alpha$ & 4 & $O$ & $\delta$ \\
5 & $~~~~~~LT~~(+)$ & $\beta$ & 5 & $RF$ & $\gamma$ \\
\hline
6 & $~~~~~~~P~~~(+)$ & $\alpha$ & 6 & $LT$ & $\gamma$ \\
7 & $LT$ & $\delta$ & 7 & $~~~~~~~P~~~(+)$ & $\alpha$ \\
8 & $~~~~~~~O~~~(+)$ & $\beta$ & & & \\
\hline
\hline
\end{tabular}
\end{center}

\end{table}

Table~\ref{tab:topmar} presents the marginal features, as measured by average RBP, that are significantly different between the CN and AD groups and between the CN and FTD groups. Combining these marginal features with the extracted PBC and NVC features as inputs, we implement the SVM algorithm for the two classification tasks in a leave-one-subject-out validation scheme. That is, one subject is left out as test data while the remaining subjects serve as the training set, and this is repeated for every subject. Then, we calculate standard performance metrics such as accuracy, sensitivity, specificity and the F1 score to assess which feature combinations best separate healthy subjects from subjects with AD and from subjects with FTD. Formulas and interpretation of these four performance metrics, and results of the classification tasks based on other machine learning techniques, such as the $k$-nearest neighbor ($k$-NN) and random forest (RF) algorithms, are also reported in detail in the Supplementary Material.

\begin{table}
\footnotesize
\caption{Classification performance of the SVM algorithm, as measured by accuracy (ACC), sensitivity (SENS), specificity (SPEC) and F1 score (F1), in the leave-one-subject-out validation scheme for the CN vs. AD and CN vs. FTD classification tasks based on PBC-based and NVC-based dependence features and their combination with the marginal features as inputs. For each classification task, different SVM input configurations are considered, including (i) the top dependence features (as measured by NVC or PBC) and (ii) combinations of the top dependence and top marginal features (as measured by average RBP). Highlighted values indicate the best performance among the PBC and NVC approaches.}
\label{tab:svm}
\begin{center}
\begin{tabular}{cccccc||cccc}

\multirow{2}{*}{\textbf{Task}} & \multirow{2}{*}{\textbf{Features}} & \multicolumn{4}{c}{\textbf{Nonlinear Vector Coherence}} & \multicolumn{4}{c}{\textbf{Pairwise Band Coherence}}\\
& & \textbf{ACC} & \textbf{SENS} & \textbf{SPEC} & \textbf{F1} & \textbf{ACC} & \textbf{SENS} & \textbf{SPEC} & \textbf{F1}\\
\hline
\hline              
\multirow{4}{*}{\textbf{CN vs. AD}} & Top 5 Dep. & 0.815 & 0.889 & 0.724 & 0.842      & 0.769 & 0.750 & 0.793 & 0.783 \\
& Top 10 Dep. & 0.862 & 0.833 & \textbf{0.897} & 0.870      & 0.631 & 0.639 & 0.621 & 0.657 \\
& Top 5 Dep. + Top 5 Mar. & \textbf{0.908} & \textbf{0.917} & \textbf{0.897} & \textbf{0.917}      & 0.800 & 0.806 & 0.793 & 0.817 \\
& Top 10 Dep. + Top 5 Mar. & 0.892 & \textbf{0.917} & 0.862 & 0.904      & 0.785 & 0.778 & 0.793 & 0.800 \\
\hline
\hline
\multirow{2}{*}{\textbf{CN vs. FTD}} & Top 5 Dep. & 0.750 & 0.759 & 0.739 & 0.772      & 0.673 & 0.793 & 0.522 & 0.730 \\
& Top 5 Dep. + Top 5 Mar. & \textbf{0.865} & \textbf{0.862} & \textbf{0.870} & \textbf{0.877}      & 0.827 & 0.828 & 0.826 & 0.842 \\
\hline
\hline
\end{tabular}
\end{center}

\end{table}

From Table~\ref{tab:svm}, we observe that the performance of the SVM algorithm is uniformly better with the NVC values as the dependence features for both CN vs. AD and CN vs. FTD classification tasks. Precisely, in classifying between healthy controls and subjects with AD, SVM obtains up to 90.8\% accuracy using the top five NVC features with the top marginal features as inputs. By contrast, accuracy based on the PBC values for the same classification task is only 80.0\% at best. This is outperformed even by the worst-case scenario for the SVM with NVC inputs. For the CN vs. FTD classification task, the dependence features extracted by the NVC measure also enables the SVM algorithm to achieve an accuracy of up to 86.5\% as compared to using PBC values which only attains 82.7\% accuracy in the best scenario.

In this paper, proposing a classification algorithm to differentiate healthy subjects from subjects in the two disease groups is addressed only as a secondary contribution. Although SVM produced reasonably good classification results, there are many other machine learning algorithms that are more sophisticated and may perform better for this goal. Rather, we highlight that the improvement in classification performance is due to our proposed NVC measure more appropriately quantifying region-to-region functional connections than the naive approach of aggregating channel-to-channel linear dependence measures such as the PBC methodology. This improvement is also observed when employing different classification methodologies such as the $k$-NN and RF algorithms with varying model specifications (see the Supplementary Material). The observed disadvantage of PBC over our proposed NVC approach likely stems from the arbitrary aggregation (by simple averaging in this case) of PBC values across possible channel pairs, which then potentially results in inadequate quantification of dependence between brain regions. Hence, our proposed NVC measure enables for a better understanding of functional connectivity alterations due to AD and FTD. This provides evidence on its practical utility in neuroscientific investigations.

\subsection{Clinical Relevance of NVC-based Functional Connectivity Patterns}\label{subchap:relevance}

\begin{figure}
	\centerline{
		\includegraphics[width=0.6\textwidth]{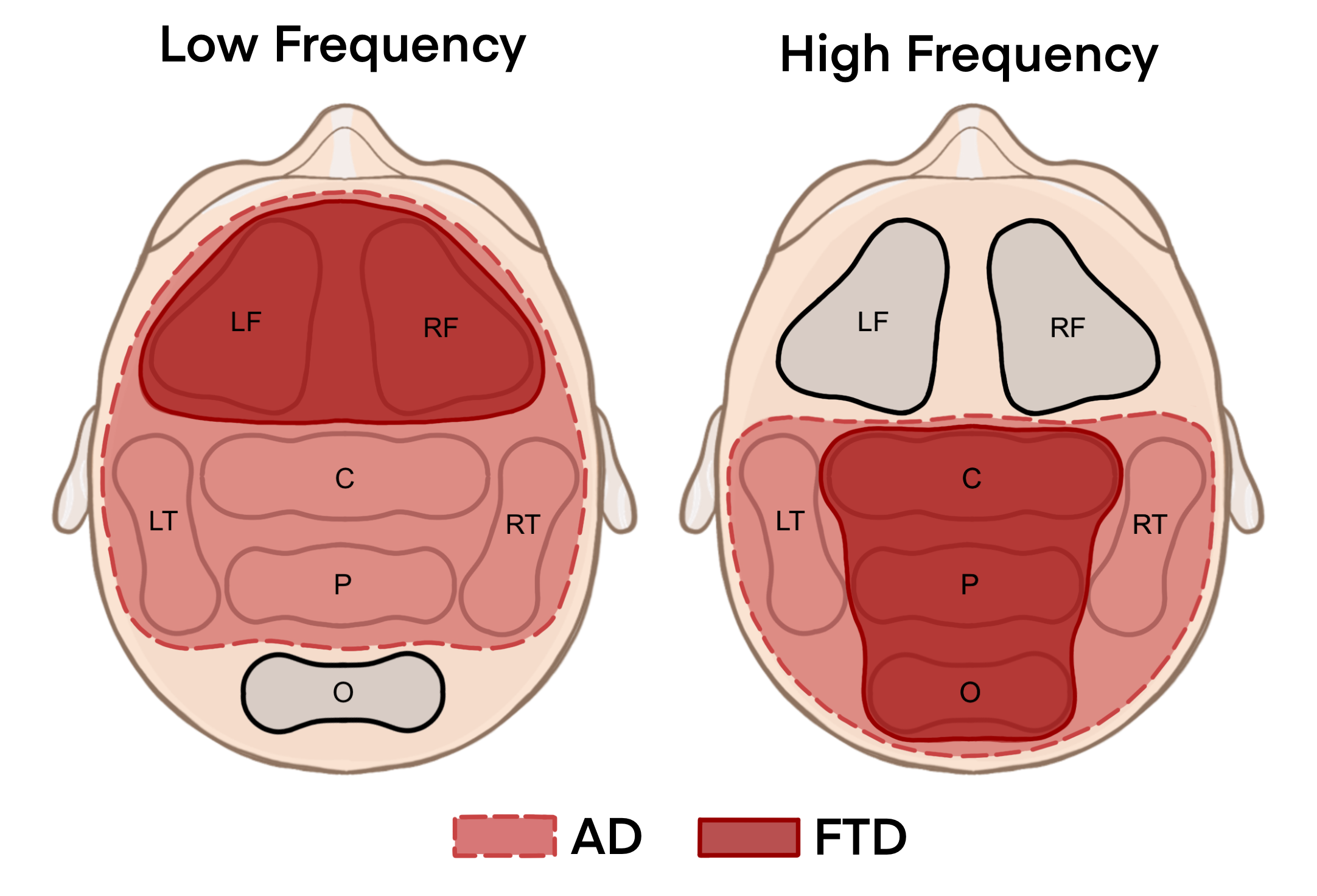}
  }
	\caption{Coverage of brain regions exhibiting functional connectivity alterations at low frequencies (left) and at high frequencies (right) among subjects with AD and FTD.}
	\label{fig:ADvsFTD}
\end{figure}

Our novel NVC measure reveals several potential biomarkers for AD and FTD. For AD, this includes the decreased functional connectivity at low frequencies (i.e., delta and theta bands) within the frontal regions, which extends to frontal-posterior interactions. Another characteristic of subject with AD is that the posterior region, involving the $LT$, $RT$, $C$, $P$ and $O$ regions, exhibit higher magnitudes of dependence at high frequencies (i.e., alpha, beta and gamma bands). Alterations in functional connectivity, similar to those identified in the AD group, are also observed among subjects with FTD, however, in a more spatially localized manner. Specifically, decreased functional connections between the left and right frontal regions at the theta band, and increased functional connections between the central and occipital regions at the beta band and between the parietal and occipital regions at the alpha, beta and gamma bands, are the most distinguishable features of the FTD group. Figure~\ref{fig:ADvsFTD} illustrates this localization phenomenon in the coverage of the brain regions that exhibit functional connectivity alterations among subjects with FTD, in comparison to the subjects with AD. Furthermore, these similar characterizations of AD and FTD may be related to the overlapping neurodysfunctions associated to the two diseases \citep{jalilianhasanpour2019functional,musa2020alzheimer}. Nonetheless, our findings are consistent with the results of \cite{zheng2025time} and may be related to the brain alterations found by \cite{chang2023quantitative} in their respective analyses of the same dataset.

As both diseases are degenerative conditions of the nervous system, the decreased functional connectivity at low frequencies in the frontal regions among subjects with AD and FTD, which are critical for executive functions, attention and working memory \citep{kolb2009fundamentals}, reflects presence of neurodegeneration that leads to neuronal loss and synaptic dysfunctions \citep{pini2016brain}. This may also be associated to the cortical thinning in the frontal lobe experienced by subjects with AD and FTD \citep{du2007different}, translating to a general cognitive decline. In addition, AD and FTD patients exhibiting increased functional connectivity at high frequencies within the posterior regions is an evidence on the brain's compensatory mechanism \citep{de2014effect,herreras2016local}. Since oscillatory activities at high frequencies are linked to active cognitive processes, e.g., focused attention \citep{pitchford2019resting}, enhanced connectivity within the posterior region suggests an attempt to compensate for the deterioration in the frontal regions \citep{zheng2025time}, which is a secondary consequence of the frontal lobe degeneration. Hence, along with the traditional amyloid beta, tau protein and fluid biomarkers for AD and FTD \citep{lashley2018molecular,swift2021fluid,klyucherev2022advances}, our NVC measure provides an alternative avenue to explore new reliable statistics-based biomarkers, which can be derived from easy-to-collect, less expensive and non-invasive modalities such as resting-state EEGs.


	\section{Conclusion and Future Directions}\label{chap:conclusion}
To address limitations of standard channel-to-channel functional brain connectivity approaches such as linear coherence, we develop the nonlinear vector coherence, a new metric for spectral dependence between multivariate time series. An advantage of our framework is that it offers a way to quantify general types of relationships between oscillations of groups of signals, and hence, gives a natural notion of region-to-region functional connectivity across different brain regions of interest in the frequency domain. One contribution of this work is a rank-based estimator for the proposed measure that (i) is fully nonparametric, and thus does not rely on restrictive model assumptions, (ii) is fast to compute, and (iii) has straightforward and biologically meaningful interpretation. Another novel contribution is our nonparametric test for independence in the frequency domain, where the empirical null distribution of the estimator is obtained via the permutations of ranks, which is then used to produce p-values. Numerical experiments demonstrate the utility of our methodology in capturing spectral dependence or independence typically observed in real-life applications, with performance that improves as the sample size increases.

In addition to the methodological novelty, the proposed NVC measure also captures novel clinical findings on the functional connectivity alterations associated with AD and FTD through the analysis of the resting-state EEG data. Results show that patients with AD and FTD exhibit weaker functional dependence at low frequencies between the frontal brain regions. By contrast, stronger functional connectivity at high frequencies are also linked to AD and FTD. Moreover, these alterations appear to be more localized for patients with FTD. Consistent with previous findings, the NVC measure provides clinically relevant characterizations of AD and FTD, which may lead to new reliable biomarkers for the two neurodegenerative conditions. We demonstrated this by showing that dependence features based on NVC lead to more accurate classification of subjects with AD or FTD v.s. healthy subjects, compared to features based on classical dependence measures.

A future direction for our work is to account for the impact of other components in the system, say a third vector time series $\boldsymbol{Z}$, that could impact or modify the relationship between $\boldsymbol{X}$ and $\boldsymbol{Y}$. One possible solution is to define a similar measure that is conditional on $\boldsymbol{Z}$, e.g., 
\begin{equation*}
    T_{\omega}(\boldsymbol{Y};\boldsymbol{X} \mid \boldsymbol{Z}) = 1 - \frac{q - \sum_{\ell=1}^q \left[\xi(\widetilde{Y}^{(\omega)}_{\ell}~; (\widetilde{\boldsymbol{X}}^{(\omega)\top},\widetilde{Y}^{(\omega)}_{1},\ldots,\widetilde{Y}^{(\omega)}_{\ell-1})^{\top} \mid \widetilde{\boldsymbol{Z}}^{(\omega)}) \right]}{q - \sum_{\ell=1}^q \left[\xi(\widetilde{Y}^{(\omega)}_{\ell}~; (\widetilde{Y}^{(\omega)}_{1},\ldots,\widetilde{Y}^{(\omega)}_{\ell-1})^{\top} \mid \widetilde{\boldsymbol{Z}}^{(\omega)}) \right]}, 
\end{equation*}
\noindent where $\widetilde{\boldsymbol{X}}^{(\omega)}$, $\widetilde{\boldsymbol{Y}}^{(\omega)}$, and $\widetilde{\boldsymbol{Z}}^{(\omega)}$ are the $\omega$-oscillations of $\boldsymbol{X}$, $\boldsymbol{Y}$ and $\boldsymbol{Z}$, as previously described in Equation~(\ref{eq:omega_XY}) and $\xi(\cdot;\cdot \mid \boldsymbol{Z})$ is now the conditional measure of \cite{azadkia2021simple}. However, further investigation is necessary to appreciate its properties and practical advantages. Other avenues for future work involves dealing with non-stationarity across different time blocks, accounting for variation across subjects within the same group, and incorporating covariate effects when estimating NVC (e.g., how NVC varies across age, clinical symptoms, and other potential factors). While this is beyond the scope of the current paper, the foundations we have established serve as a strong starting point for developing distribution-free spectral dependence measures between vector time series and contributes to a better understanding of functional connectivity across brain regions.

    \clearpage
    \newpage
    \bibliographystyle{biom} 
    \bibliography{7_references}
    \label{lastpage}

\end{document}